%% file: paper.tex
\documentclass[prc,showpacs,preprintnumbers,amsmath,superscriptaddress,floatfix,nofootinbib]{revtex4-1}

\usepackage[toc,page]{appendix}
\usepackage{bm}
\usepackage{color}
\usepackage[dvipsnames]{xcolor}
\usepackage{graphics}
\usepackage{rotating}

\usepackage{graphicx}
\DeclareGraphicsExtensions{.pdf,.png,.jpg,.ps}

\newcommand{\pion}{{\pi}}
\newcommand{\kaon}{{K}}
\newcommand{\kaonb}{{\bar{K}}}
\newcommand{\kaonS}{{K^*}}
\newcommand{\kaonSb}{{\bar{K}^*}}

\newcommand{\Nucleon}{{N}}

\newcommand{\SigmaS}{{\Sigma^*}}

\newcommand{\Cascada}{{\Xi}}

\newcommand{\CascadaS}{{\Xi^*}}

\newcommand{\SU}{{\rm SU}}
\newcommand{\U}{{\rm U}}

\newcommand{\ignore}[1]{} 

\usepackage{array}
\newcolumntype{L}[1]{>{\raggedright\let\newline\\\arraybackslash\hspace{0pt}}m{#1}}
\newcolumntype{C}[1]{>{\centering\let\newline\\\arraybackslash\hspace{0pt}}m{#1}}
\newcolumntype{R}[1]{>{\raggedleft\let\newline\\\arraybackslash\hspace{0pt}}m{#1}}

\begin{document}

\title{Compositeness of the strange, charm and beauty  odd parity $\Lambda$ states}

\author{C. Garcia-Recio}
\affiliation{Departamento de F\'isica At\'omica, Molecular y Nuclear, and Instituto Carlos I de F\'isica
Te\'orica y Computacional, Universidad de Granada, E-18071 Granada, Spain}
\author{C. Hidalgo-Duque}
\affiliation{Instituto de F\'isica Corpuscular (IFIC),
             Centro Mixto CSIC-Universidad de Valencia,
             Institutos de Investigaci\'on de Paterna,
             Aptd. 22085, E-46071 Valencia, Spain}
\author{J. Nieves}
\affiliation{Instituto de F\'isica Corpuscular (IFIC),
             Centro Mixto CSIC-Universidad de Valencia,
             Institutos de Investigaci\'on de Paterna,
             Aptd. 22085, E-46071 Valencia, Spain}
\author{L.L. Salcedo}
\affiliation{Departamento de F\'isica At\'omica, Molecular y Nuclear, and Instituto Carlos I de F\'isica
Te\'orica y Computacional, Universidad de Granada, E-18071 Granada, Spain}
\author{L. Tolos}
\affiliation{Instituto de Ciencias del Espacio (IEEC/CSIC), Campus Universitat Aut\`onoma de
Barcelona, Carrer de Can Magrans, s/n, 08193 Cerdanyola del Vall\`es, Spain}
\affiliation{Frankfurt Institute for Advanced Studies. Johann Wolfgang Goethe University, Ruth-Moufang-Str. 1,
60438 Frankfurt am Main, Germany }

\begin{abstract}
   We study the dependence on the quark mass of the compositeness of
  the lowest-lying odd parity hyperon  states. Thus, we pay attention
  to    $\Lambda-$like states in the strange, charm and
  beauty, sectors which are dynamically generated using a unitarized meson-baryon model. In the strange
  sector we use an $\SU(6)$ extension of the Weinberg-Tomozawa meson-baryon
  interaction, and we further implement the heavy-quark spin symmetry to
  construct the meson-baryon interaction when charmed or beauty hadrons are
  involved.  In the three examined flavor sectors, we obtain two $J^P=1/2^-$  and one $J^P=3/2^-$
  $\Lambda$ states. We find that
  the $\Lambda$ states which are bound states (the three $\Lambda_b$) or
  narrow resonances (one $\Lambda(1405)$ and one $\Lambda_c(2595)$) are well
  described as molecular states composed of $s$-wave meson-baryon pairs. The
   $\frac{1}{2}^-$ wide $\Lambda(1405)$ and $\Lambda_c(2595)$  as well as the
  $\frac{3}{2}^-$ $\Lambda(1520)$ and $\Lambda_c(2625)$ states display smaller
  compositeness and so they would require new mechanisms, such as $d$-wave
  interactions.
\end{abstract}

\date{\today}

\maketitle

\vspace{1cm}

\section{Introduction}
\label{sect::introduction}

One of the chief theoretical efforts in hadron physics is to
understand the nature of hadrons, whether they can be primarily
explained within the quark-model picture as multiquark states or
mainly qualify as dynamically generated states via hadron-hadron
scattering processes. In particular, in the last years there has been
a growing interest in the properties of strange and charmed baryons in
connection with many experiments such as the on-going CLEO
\cite{cleo}, Belle \cite{belle}, BES \cite{bes}, BaBar \cite{babar} as
well as the planned PANDA \cite{panda} or the J-PARC upgrade
\cite{jparc}. Also, the LHCb Collaboration at CERN has been exploring
in the recent years an almost {\em terra incognita} in the spectroscopy of
baryons with the beauty degree of freedom. Results on beauty baryonic
states, such as the $\Lambda_b$ excited states \cite{Aaij:2012da},
have been reported, stimulating the theoretical work to understand the
properties of the newly-discovered states.

Recent approaches based on unitarized coupled-channels methods have
proven to be very successful in describing the existing experimental
data in the charmed
\cite{Tolos:2004yg,Tolos:2005ft,Lutz:2003jw,Lutz:2005ip%
,Hofmann:2005sw,Hofmann:2006qx,korpa-lutz,Mizutani:2006vq%
,Tolos:2007vh,JimenezTejero:2009vq,JimenezTejero:2011fc,Haidenbauer:2007jq%
,Haidenbauer:2008ff,Haidenbauer:2010ch,Wu:2010vk,Wu:2010jy%
,Oset2012,Liang:2014kra}
and beauty baryonic sectors~\cite{Wu:2010rv,Liang:2014eba}.  Most of
these models emerge as the theoretical effort extends from the strange
to charmed and beauty sectors, partially motivated by the parallelism
between the $\Lambda(1405)$ and the $\Lambda_c(2595)$ as well as the
$\Lambda_b(5912)$ states. Of special importance are the symmetries
that are implemented in the hadronic models. While chiral symmetry
should be implemented in the strangeness sector, heavy-quark spin
symmetry (HQSS) \cite{IW89,Ne94,MW00} appears naturally as we deal
with systems that include charmed and beauty degrees of freedom
\cite{GarciaRecio:2008dp,Gamermann:2010zz,Romanets:2012hm,GarciaRecio:2012db%
,Garcia-Recio:2013gaa,Tolos:2013gta,Tolos:2009nn,GarciaRecio:2010vt,GarciaRecio:2011xt,Xiao:2013yca,Ozpineci:2013zas}.

The use of the effective models combined with unitarity constraints in coupled
channels allows to explain many baryons in terms of meson-baryon interactions,
interpreting them as composite or dynamically-generated states. The ultimate
goal is to determine the degree of ``compositeness" and the ``genuine"
contributions of the given state. The formalism was developed by Weinberg in
Ref. \cite{Weinberg:1962hj}, and later applied to the deuteron in
\cite{Weinberg:1965zz}, showing that the deuteron can be fully understood as
a proton-neutron bound state. More recent works have extended this analysis from
bound states to resonances and from $s$-wave to higher partial waves
\cite{Hanhart:2010wh,Baru:2003qq,Gamermann:2009uq,YamagataSekihara:2010pj%
  ,Cleven:2011gp,Aceti:2012dd,Xiao:2012vv,Aceti:2014ala}. The theoretical
aspects have been further discussed in
\cite{Hyodo:2011qc,Nagahiro:2014mba}. 

The present paper is focused on the analysis of the compositeness of
the lowest-lying $J^P=1/2^-$ and $J^P=3/2^-$ $\Lambda$ states going
from the strange to the beauty sectors.\footnote{From here on we shall
  use $\Lambda$ to indistinctly denote the $\Lambda$, $\Lambda_c$ and
  $\Lambda_b$ states} The aim is to shed some light on the nature of
newly discovered excited $\Lambda_c$ and $\Lambda_b$ by exploiting the
similarities with the strange $\Lambda$ states. We also address any
existing regularity in the quark mass dependence of the compositeness of these
excited baryons. 

There exist previous studies in the strange sector. The
$\Lambda(1520)$ has been recently discussed in
Ref.~\cite{Aceti:2014wka}, while the compositeness 
and elementariness
of the two $\Lambda(1405)$ states have been evaluated within the chiral
unitarity model (Goldstone boson-baryon chiral perturbation potential
used as a kernel of a Bethe-Salpeter equation) at leading order (Weinberg-Tomozawa
interaction) in Ref.~\cite{Sekihara:2012xp} and incorporating next-to-leading chiral
corrections~\cite{Sekihara:2014kya}. While the results reported in
Refs.~\cite{Sekihara:2014kya, Sekihara:2012xp} are in qualitative
agreement with those obtained in this work, there are some appreciable
differences between our approach and that followed in
\cite{Aceti:2014wka} for the $\Lambda(1520)$ resonance, namely the
consideration of $d$-wave interactions, which effects will be further
investigated.  

The structure of this
work is as follows. In Section \ref{sect::formalism} we summarize the
model we are using to describe meson-baryon interactions and present
the dynamically-generated $\Lambda$ resonances in the strange, charm
and beauty sectors.  In Sect.~\ref{sect::compositeness} we use the
generalized Weinberg's sum rule to estimate the importance of the
different channels in the generated $\Lambda$ states. With this
analysis, we give an educated guess of the compositeness of these
states.  Finally, some conclusions are addressed in
Sect.~\ref{sect::conclusions}.   In Appendix \ref{app:Gp} we present the analytical expression of compositeness on the first and second Riemann sheets, whereas in Appendix \ref{app:sumrule} the compositeness rule is derived.

\section{Dynamically-generated  strange, charm and beauty $\Lambda$ states}
\label{sect::formalism}

In order to analyze the compositeness of the $\Lambda$ (isoscalar) states in
the strange, charm or beauty sectors, we start by summarizing the meson-baryon
model used in the different sectors.

On one hand, for meson-baryon interactions involving strange hadrons, we
consider the extension of the Weinberg-Tomozawa (WT) interaction to SU(6) of
Refs.~\cite{GarciaRecio:2005hy, GarciaRecio:2006wb, Toki:2007ab,
  GarciaRecio:2010ki,Gamermann:2011mq,Garcia-Recio:2013uva}.  This is an
$s$-wave contact interaction.  This model assumes that the light quark--light
quark interaction is approximately spin independent as well as SU(3) 
flavor symmetric, thus, treating the six states of a light quark as equivalent.
This approach has shown a reasonable semi-qualitative outcome for different
excited baryonic states as compared to the experimental data
\cite{Gamermann:2011mq}.

On the other hand, in the heavy sectors of charm and beauty, we extend the
model by making use of the HQSS to construct meson-baryon interactions
involving heavy hadrons. Whereas HQSS connects vector and pseudoscalar mesons
containing heavy quarks as all type of spin interactions vanish for infinitely
massive quarks, chiral symmetry fixes the lowest order interaction between
Goldstone bosons and other hadrons in a model independent way via the WT
interaction. Then, it is appealing to construct a model for four flavors
including all basic hadrons (pseudoscalar and vector mesons, and $1/2^+$ and
$3/2^+$ baryons) which reduces to the WT interaction in the sector where
Goldstone bosons are involved and which incorporates HQSS in the sector where
heavy quarks participate \cite{GarciaRecio:2008dp, Gamermann:2010zz,
  Romanets:2012hm,GarciaRecio:2012db,Garcia-Recio:2013gaa}. This
SU(6)$\times$HQSS model is justified in view of the reasonable semi-qualitative
outcome of the SU(6) extension \cite{Gamermann:2011mq} and on a formal
plausibleness on how the SU(4) WT interaction in the heavy pseudoscalar
meson-baryon sectors comes out in the vector-meson exchange picture.
\input{masses.tex}

The extended WT meson-baryon interaction, in the coupled meson-baryon
basis with total heavy content  (charm $C$ /beauty $B$) $H$, strangeness $S$,
isospin $I$ and spin $J$, is given by
\begin{equation}
V_{ij}^{HSIJ} =
D_{ij}^{HSIJ}\,\frac{2\sqrt{s}-M_i-M_j}{4f_if_j} 
\sqrt{\frac{E_i+M_i}{2M_i}}\sqrt{\frac{E_j+M_j}{2M_j}}, 
\label{eq:pot}
\end{equation}
where $\sqrt{s}$ is the center of mass (C.M.) energy of the system; $E_i$ and
$M_i$ are, respectively, the C.M. on-shell energy and mass of the baryon in the channel
$i$; and $f_i$ is the decay constant of the meson in the $i$-channel.
Symmetry breaking effects are introduced by using physical masses and decay
constants.  The masses and decay constants used in this work are shown in
Table \ref{tab:masses}. The masses shown correspond to the
arithmetic mean of the different isospin partners.

The $D_{ij}^{HSIJ}$ are the matrix elements coming from the group structure of
the extended WT interaction \cite{Garcia-Recio:2013gaa}. The matrix elements
required for the $\Lambda(1520)$ sector, with quantum numbers
$C=0,~B=0,~S=-1,~I=0, ~J^P=3/2^-$, can be found in Eq.~(45) of
Ref.~\cite{Toki:2007ab}. Those for the $\Lambda_c(2595)$, with
$C=1,~B=0,~S=0,~I=0,~J^P=1/2^-$ and $\Lambda_c(2625)$, with $J^P=3/2^-$, can
be found  in Tables XV and XVIII of Ref.~\cite{GarciaRecio:2008dp},
respectively. The same coefficients apply to the bottom case. Finally, the
matrix elements for the $\Lambda(1405)$ sector, with quantum numbers
$C=0,~B=0,~S=-1,~I=0,~J^P=1/2^-$, can be extracted following the directions of
Appendix A of Ref.~\cite{Gamermann:2011mq} and the conventions in
\cite{GarciaRecio:2010vf}. For convenience these matrix elements are explicitly
displayed in Table~\ref{tab:00m002}.  \input{t00m112.tex}

In order to obtain the unitarized $T$-matrix, we solve  the
on-shell factorized form of the Bethe-Salpeter equation using the matrix $V^{HSIJ}$ as kernel
\begin{equation}
T^{HSIJ}=(1-V^{HSIJ}G^{HSIJ})^{-1}V^{HSIJ}
\label{eq:bse} \ ,
\end{equation}
where $G^{HSIJ}$ is a diagonal matrix containing the meson-baryon propagator
in each channel. Explicitly,
\begin{equation}
\label{eq::funcion_loop}
G_i^{HSIJ}(\sqrt{s},m_i,M_i) = i ~2 M_{i} \int \frac{d^{4}q}
{\left( 2\pi \right)^{4}} \frac{1}{q^{2} - m_i^2} \frac{1}{\left(P-q\right)^{2} - M_i^2},
\end{equation}
being $M_i(m_i)$ the baryon (meson) mass of the channel $i$ and $P^\mu$ the
total four-momentum, which in the CM frame is given by $P_{\mathrm{CM}}^\mu =
(\sqrt{s},\mathbf{0})$.  The loop function is explicitly given in
Ref. \cite{Nieves:2001wt} and in Appendix \ref{app:Gp}.

When the meson and/or the baryon in the intermediate state is not a stable
particle, we convolute the meson-baryon propagator (loop) with the
corresponding hadronic spectral function, as done in
Refs. \cite{Roca:2006sz,Gamermann:2007fi,Gamermann:2010zz}). Thus, in this
case, the loop function $G$ is substituted by $\hat{G}$, which is defined as
the convolution of the loop function $G$ with the spectral function of this
intermediate resonant state (R),
\begin{eqnarray}
 \hat{G}^{HSIJ}(\sqrt{s},m,M_R,\Gamma_R) =  
\frac{1}{N} \int_{(M_R - 2\Gamma_R)^2}^{(M_R + 2\Gamma_R)^2} d\hat{M}^{2} 
\left(- \frac{1}{\pi}\right) {\rm{Im}} 
\left(\frac{1}{\hat{M}^2 - M_R^2 
+ i M_R \Gamma_R}\right) G^{HSIJ}(\sqrt{s},m,\hat{M}),
\label{eq:convolution}
\end{eqnarray}
being $N$ a normalization factor that reads,
\begin{equation}
N =  \int_{(M_R - 2\Gamma_R)^2}^{(M_R + 2\Gamma_R)^2} d\hat{M}^{2} 
\left( -\frac{1}{\pi}\right) {\rm{Im}}
\left(\frac{1}{\hat{M}^2 - M_R^2 + i M_R \Gamma_R}\right) .
\end{equation}

The meson-baryon propagator is logarithmically ultraviolet divergent, thus, it
needs to be renormalized. This has been done by a subtraction point
regularization such that
\begin{equation}
G_{ii}^{HSIJ} (\sqrt{s})=0 \quad\text{at~~} \sqrt{s}=
\mu^{HSI},
\label{eq:musi}
\end{equation}
with 
\begin{equation}
\mu^{HSI} = \sqrt{\alpha}\sqrt{ m_{\rm{th}}^2+M_{\rm{th}}^2 }
,
\label{eq:mu}
\end{equation}
where $m_{\rm{th}}$ and $M_{\rm{th}}$, are, respectively, the masses of the
meson and baryon producing the lowest threshold (minimal value of
$m_{\rm{th}}+M_{\rm{th}}$) for each $HSI$ sector, independent of the angular
momentum $J$, and $\alpha=1$. This renormalization scheme was first proposed
in Refs.~\cite{Hofmann:2005sw,Hofmann:2006qx} and it was successfully used in
Refs.~\cite{GarciaRecio:2003ks,GarciaRecio:2008dp,Gamermann:2010zz,Romanets:2012hm,Garcia-Recio:2013gaa,GarciaRecio:2012db}. A recent
discussion on the regularization method can be found in
Ref.~\cite{Hyodo:2008xr}. The overall results obtained by the above choice of
subtraction point is similar to the observed spectrum of low-lying hadronic
resonances. A more precise agreement can be achieved by suitably shifting the
subtraction point. To do so one can choose a value of the parameter $\alpha$
different from unity \cite{GarciaRecio:2008dp,GarciaRecio:2012db}. Note that,
other than this, the model has no free parameters.

The dynamically-generated baryon states appear as poles of the scattering
amplitudes on the complex energy $\sqrt{s}$ plane. The poles of the scattering
amplitude on the first Riemann sheet that appear on the real axis below
threshold are interpreted as {\it bound states}. The poles that are found on
the second Riemann sheet below the real axis and above threshold are
identified with {\it resonances}.\footnote{For convenience we will often use
  the word resonance for all molecular states discussed in this work, whether
  they are bound states or proper resonances.} The mass and the width of the
state can be found from the position of the pole on the complex energy
plane. Close to the pole, the scattering amplitude behaves as
\begin{equation}
T^{HSIJ}_{ij} (s) \approx \frac{g_i\,g_j}{\sqrt{s}-\sqrt{s_R}} \,.
\label{Tfit}
\end{equation} %
The mass $M_R$ and width $\Gamma_R$ of the state result from $\sqrt{s_R} = M_R -
\rm{i}\, \Gamma_R/2$, while $g_j $ (complex in general) is the coupling of the
state to the $j$-channel.

The calculated positions and widths of the lowest-lying $\Lambda$ states in
the strange, charm and beauty sectors together with their couplings, $g_i$, to
the different meson-baryon channels are shown in Tables \ref{tab:La1520},
\ref{tab:LaC2625} and \ref{tab:LaB5920}.  In the case we want to refer to a
specific flavor we will write $\Lambda_s$, $\Lambda_c$ or $\Lambda_b$.  For
each flavor $f = s,c,b$, the resonances $\Lambda_f$ are ordered by closeness
to the $\pi\Sigma_f$ threshold, and they are displayed in this sequence in the
Tables.

We use the convoluted meson-baryon propagator for the non-stable intermediate
particles (namely, $\rho$, $K^*$ and $\bar{K}^*$ mesons and $\Sigma^*$
baryon) in the study of the
strange sector for the $\Lambda(1405)$ and $\Lambda(1520)$ resonances, in a
similar manner as done in Ref.~\cite{Gamermann:2011mq}. In
Ref.~\cite{Romanets:2012hm}, it was reported that the convolution did not
affect the dynamically generated $\Lambda_c$ states in a substantial manner as
the dominant convoluted meson-baryon channels were far from the position of
the heavy $\Lambda_c$ states.

In view of their mass position and dominant couplings, we assign these states
to the experimental strange [$\Lambda(1405)$, $\Lambda(1520)$], charmed
[$\Lambda_c(2595$, $\Lambda_c(2625)$] and beauty [$\Lambda_b(5912)$,
  $\Lambda_b(5920)$] states, similarly to
Refs.~\cite{GarciaRecio:2008dp,Gamermann:2011mq%
  ,Romanets:2012hm,GarciaRecio:2012db}. Note, however, that in
Refs.~\cite{GarciaRecio:2008dp,GarciaRecio:2012db} the subtraction point was
slightly modified in order to fix the position of the dynamically-generated
states to the experimental predictions of the $\Lambda_c(2595)$ and
$\Lambda_b(5912)$, respectively.

Three $\Lambda$ states are obtained in each of the flavor sectors, two of them
with $J^P=1/2^-$ and one with $J^P=3/2^-$. The well-known two-pole pattern of
the $\Lambda(1405)$ \cite{Jido:2003cb,GarciaRecio:2002td,GarciaRecio:2003ks}
is reproduced for the $\Lambda_c(2595)$ and $\Lambda_b(5912)$. Indeed, for
$J^P=1/2^-$ we find a state that strongly couples to $N M$ and $N M^*$
channels, with $(M,M^*)= (\bar K, \bar K^*)$, $(D, D^*)$ or $(\bar B, \bar
B^*)$ for strange, charm or beauty sectors, respectively.  The $\bar{K} N$
dominance in the $\Lambda(1405)$ has been got some support from  lattice QCD
calculations \cite{Hall:2014uca}. In addition, a second state $1/2^-$ coupling
to $B \pi$, with $B=\Sigma$, $\Sigma_c$ or $\Sigma_b$ is also seen for
strangeness, charm or beauty, respectively. (In what follows we simply refer
to these two $\Lambda(\frac{1}{2}^-)$ states as ``first'' and ``second''
state, respectively). On the other hand, the $J^P=3/2^-$ states
$\Lambda_c(2625)$ and $\Lambda_b(5920)$ are the counterparts in the charm and
beauty sectors of the $\Lambda(1520)$.
 
In Refs.~\cite{GarciaRecio:2008dp,Gamermann:2011mq,Romanets:2012hm%
  ,GarciaRecio:2012db}, the {\em coupling constants} were interpreted as a
measure of the importance of a channel in order to determine the molecular
nature of the state. For instance, the $\Lambda(1405)$ state close to the
scattering line would be a mixture of $\bar{K}N$ and $\bar{K}^*N$ states,
while the second $\Lambda(1405)$ state, with a very large decay width, would
be mainly a $\pi\Sigma$ state. In the next section, we argue that the coupling
constants, though useful, are not sufficient to describe the nature of a
resonance. Thus, further analyses of the nature and, hence, of the {\em
  compositeness} of the $\Lambda$ states are required.

\section{Compositeness of the $\Lambda$ states}
\label{sect::compositeness}

In Ref.~\cite{Weinberg:1965zz} Weinberg analyzed the nature of the deuteron
and found that this particle is best described as composed of a proton and a
neutron, rather than a genuine dibaryon. More recently, the issue of
compositeness was addressed in
Ref.~\cite{Hanhart:2010wh,Baru:2003qq,Cleven:2011gp} for $s$-waves and small
binding energies. An extension to larger binding energies in coupled-channel
dynamics was undertaken in Ref.~\cite{Gamermann:2009uq} for bound states and
in Refs.~\cite{YamagataSekihara:2010pj,Xiao:2012vv,Aceti:2012dd,Aceti:2014ala}
for resonances.  In this section we summarize the formalism and the
conclusions derived in Ref. \cite{Aceti:2014ala} for the interpretation of the
Weinberg's sum rule and its generalization to resonances.

In the unitarized setting the sum rule follows from the
identity~\cite{Hyodo:2011qc, Sekihara:2012xp,Hyodo:2013nka,Sekihara:2014kya}:
\begin{equation}
-1 =  \sum_{i,j} g_i  g_j \left.\left( \delta_{ij} \frac{\partial
 G_{i}(\sqrt{s})}{\partial \sqrt{s}}
+
  G_i(\sqrt{s}) \frac{\partial V_{ij}(\sqrt{s})}
{\partial \sqrt{s}} G_j(\sqrt{s})  \right)\right|_{\sqrt{s}=\sqrt{s_R}} 
\,.
\label{eq::sumrule_energy_dependent}
\end{equation}
This relation is derived in Appendix \ref{app:sumrule}. It holds for bounds
states and resonances, as well as energy dependent or energy independent
interactions.

The use of the definitions
\begin{equation}
X_i = - {\rm{Re}} \left(g_{i}^{2} \left. \frac{d G_{i}}{d\sqrt{s}} 
\right|_{\sqrt{s_R}}\right)
,\qquad
Z =  - {\rm{Re}} \sum_{i,j} g_i    g_j  \left( \left. 
  G_i \frac{\partial V_{ij}}
{\partial \sqrt{s}} G_j  \right)\right|_{\sqrt{s_R}}
\label{eq:sumruledef}
\end{equation}
provides the sum rule
\begin{equation}
1 = Z + \sum_i X_i
\,.
\label{eq:sumrule}
\end{equation}
For bound states the extraction of the real part in Eq.~(\ref{eq:sumruledef}) is
redundant since the quantities involved are already real. The expression of
$X_i$ involves the derivative of the loop function. The analytical expression
of this function on the first and second Riemann sheets is made explicit in
Appendix \ref{app:Gp}.

As follows from the analysis in \cite{Aceti:2014ala}, for bound states, the
quantity $X_i$ is related to the probability of finding the state in the
channel $i$. For resonances, $X_i$ is still related to the squared wave
function of the channel $i$, in a phase prescription that
automatically renders the wave function real for bound states, and so it can be used as a measure of the weight of
that meson-baryon channel in the composition of the resonant state.

The quantity $\sum_i X_i = 1-Z$ represents the {\em compositeness} of the
hadronic state in terms of all the considered channels, and $Z$ is referred to
as its {\em elementariness}. A non vanishing $Z$ takes into account that
ultimately the model is an effective one. The energy dependent interaction
effectively accounts for other possible interaction mechanisms not explicitly
included in the $s$-wave meson-baryon description.  These could be other
hadron-hadron interactions, or even genuine negative-parity baryonic
components not of the molecular type (hence the appellative
elementariness). Thus, a small value of $Z$ indicates that the state is well
described by the contributions explicitly considered, namely, $s$-wave
meson-baryon channels. Conversely, a larger value of $Z$ indicates that, for
that state, significant pieces of information are missing in the model, and
this information is being included through an effective interaction,
to the extent that the experimental hadronic properties are reproduced by the
model.

The results we obtain for the compositeness weights, $X_i$, and
aggregated compositeness $1-Z$ of the various $\Lambda$ states are
displayed in Tables \ref{tab:La1520}, \ref{tab:LaC2625} and
\ref{tab:LaB5920}, for the default value $\alpha=1$ and also for
another phenomenological choice of the subtraction point, so that the
experimental masses are better reproduced. As mentioned in the
introduction, the results reported in Refs.~\cite{Sekihara:2014kya,
  Sekihara:2012xp} are in qualitative agreement with those presented
in Table~\ref{tab:La1520} for the $\Lambda(1405)$ states. In what
follows we draw some conclusions with regards to the nature of the
$\Lambda$ states and its variation with the quark mass can be
extracted from the numbers.

First, the contribution of each meson-baryon channel to the dynamical
generation of a state is determined not only by the value of the coupling
constant but also depends on the closeness of meson-baryon channel to the
state. For instance, the $\bar{K}N$ and $\bar{K}^* N$ channels have similar
couplings to the  first pole of $\Lambda(1405)$  but their contribution to
the compositeness is quite different due to their different thresholds,
relative to the mass of the state.

Second, the neglected contributions can be measured by means of the
elementariness. Indeed, we observe that those $\Lambda$ poles close to the
scattering line are well described as molecular states through the $s$-wave
meson-baryon channels considered, while wider states need the consideration of
other contributions, such as multi-hadron scattering. This is clearly manifest
for the $J^P=3/2^-$ states $\Lambda(1520)$ and $\Lambda_c(2625)$. There is,
however, not a strict correlation between the value of the width and the
elementariness. The $1/2^-$ states have a larger compositeness than their
$3/2^-$ counterparts.

Third, taking the natural identification between different $\Lambda$ states
for different flavors, one observes that as a rule, the heavier the flavor the
larger the compositeness of the resonance. For instance, the $\Lambda(1520)$,
$\Lambda_c(2625)$ and $\Lambda_b(5920)$ states have $1-Z=0.27$, $0.37$, and
$0.82$, respectively (for $\alpha=1$).


In the tables we primarily display results for the default value $\alpha=1$,
even though this choice of subtraction point does not reproduce the empirical
masses of the resonances in detail. We also display results with $\alpha$
suitably fitted in each case so that empirical masses of the resonances are
reproduced. For the sake of definiteness, an equal mass for the two $1/2^-$
$\Lambda$ states of each flavor has been adopted. The purpose of doing this is
not to achieve a precise description of the resonance, but rather to see to
what extent the subtraction point and the resonance position are relevant for
the compositeness discussion. We can see that no substantial modifications in
the weights $X_i$ take place in the charm and beauty cases, and the same holds
for the first $\Lambda(1405)$ state. The change is somewhat larger for the
second $\Lambda(1405)$ state and for the $\Lambda(1520)$ resonance. For these
two resonances, the change required in the subtraction points is also sizable.

In order to understand these features, one can observe that the three first
$\Lambda$ states, namely, $\Lambda(1405)$, $\Lambda_c(2595)$ and
$\Lambda_b(5912)$, have sizable weights ($X_i$) in the nucleon-pseudoscalar
channel, $\bar{K}N$, $DN$ and $\bar{B}N$, respectively, while the weights of
the $\Sigma$-pseudoscalar lightest channels, $\pi \Sigma$, $\pi \Sigma_c$ and
$\pi \Sigma_b$, are much smaller or even negligible in the bottom case. The
couplings ($g_i$) to these two types of channels follow a similar trend, and
this explains the small widths of these resonances.  In fact, for the
$\Lambda(1405)$, the $\bar{K}N$ channel is dominant as regards to
compositeness (although the coupling to $\bar{K}^*N$ is also large). For the
$\Lambda_c(2595)$ and $\Lambda_b(5912)$, the weight of $DN$ and $\bar{B}N$ is
important but competes with $D^*N$ and $\bar{B}^*N$. For the charm (bottom)
sector  this was also found in  \cite{GarciaRecio:2008dp,
  Romanets:2012hm} (\cite{GarciaRecio:2012db}) and in \cite{Liang:2014kra}. Likely, this is a
consequence of the similar roles played by vector and pseudoscalar heavy
mesons ($D$ and $D^*$ or $\bar{B}$ and $\bar{B}^*$) due to heavy quark
symmetry. The fact that these first $\Lambda(1405)$, $\Lambda_c(2595)$ and
$\Lambda_b(5912)$ poles have compositeness $1-Z$ close to unity indicates that
the present model, with $s$-wave meson-baryon including pseudoscalar and
vector mesons, gives a fair description of these resonances.

Likewise, the compositeness is large in the case of the second
$\Lambda_b(5912)$ and the $\Lambda_b(5920)$, suggesting that the model
is also fairly complete for these two resonances.

Small values of $1-Z$, below $0.5$, are found for the second $\Lambda(1405)$
and the second $\Lambda_c(2595)$ in the $1/2^-$ sector, as well as the
$\Lambda(1520)$ and $\Lambda_c(2625)$ in the $3/2^-$ sector. A conspicuous
difference between the first and second $\Lambda(1/2^-)$ resonances is that the
latter states strongly couple to the lightest channel $\pi\Sigma$ or
$\pi\Sigma_c$, and these channel largely saturate their compositeness $1-Z$.
The same applies to the $3/2^-$ $\Lambda$ states, this time with $\pi\Sigma^*$
or $\pi\Sigma_c^*$ channels.  As a consequence, these four resonances have a
sizable width. Related to this, the available phase-space of the meson-baryon
pair allows mechanisms involving higher partial waves (beyond $s$-wave) to
play a role in the composition of the resonance.  These missing mechanisms
would be accounted for by the larger values of $Z$ displayed by these four
resonances.

\begin{figure*}
\begin{center}
\includegraphics[width=0.8\textwidth]{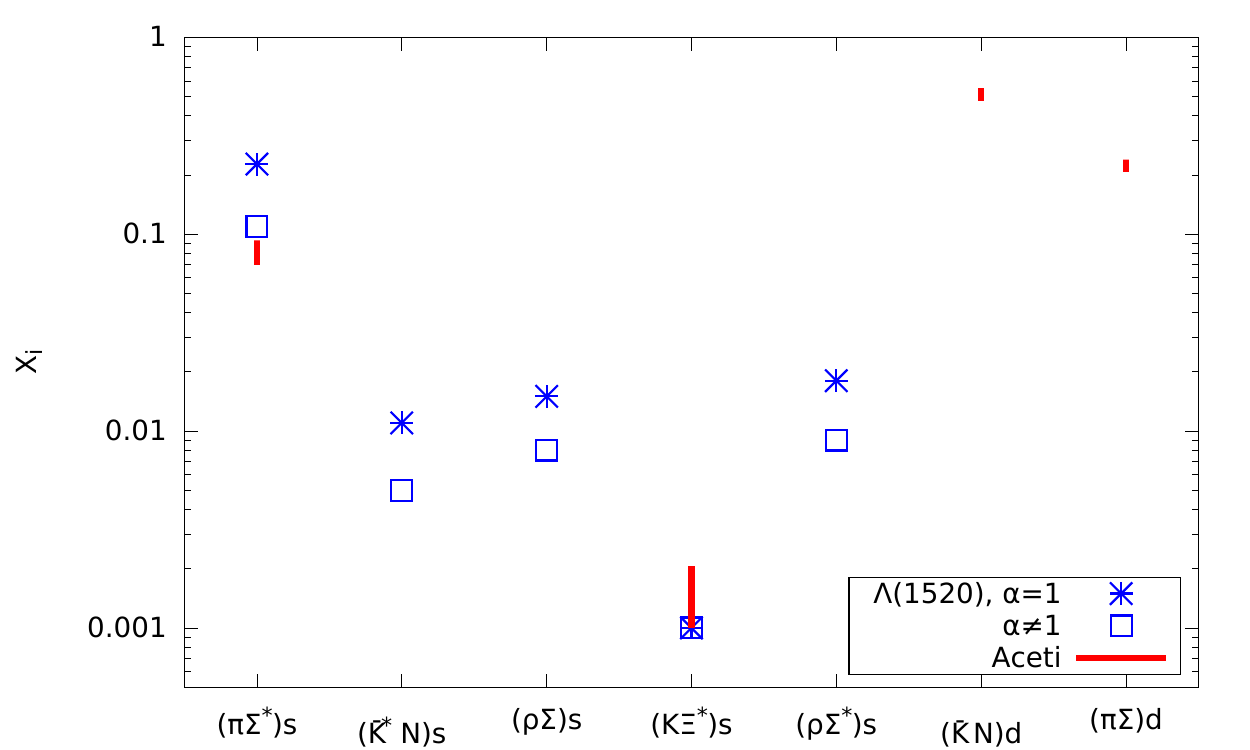}
\caption{\small Weights $X_i$ of the main channels contributing to the
  composition of the $\Lambda(1520)$.  Our results (in blue) are represented
  by stars for $\alpha=1$, and by squares when the subtraction point is
  modified to bring the mass of the resonance to its experimental value.  The
  vertical lines in red indicate the weights obtained in \cite{Aceti:2014wka}
  for the two $s$-wave and two $d$-wave channels considered there using
  various sets of fitting parameters.  }
\label{fig:1}
\end{center}
\end{figure*}

Within the molecular approach, the first missing interaction mechanism is
expected to come from $d$-wave interactions. These type of interactions have
been considered in Ref.~\cite{Aceti:2014wka} for the $\Lambda(1520)$. The
specific channels considered there are $\pi\Sigma^*$ and $K\Xi^*$ in $s$-wave,
and $\bar{K}N$ and $\pi\Sigma$ in $d$-wave. Further, the interaction is
modeled as to reproduce $\bar{K}N$ scattering data, and several fits
consistent with the experimental mass of the $\Lambda(1520)$ are
presented. That calculation suggests that $d$-wave components play an
important role in the structure of the $\Lambda(1520)$.  In Fig.~\ref{fig:1}
we display a comparison between our results and those in \cite{Aceti:2014wka}
for the weights of each channel.  The vertical lines interpolate between the
different values given in that work for different fits. While we
have not included higher partial waves in our interaction, we find that the
weights of the $s$-wave channels included in \cite{Aceti:2014wka} are
qualitatively similar in both calculations and the agreement improves as the
position of the pole is moved to its experimental value by a change of
subtraction point.  It can also be seen that other $s$-wave channels are more
relevant than $K\Xi^*$, namely, $\bar{K}^* N$, $\rho \Sigma$ and $\rho
\Sigma^*$, although $\pi\Sigma^*$ is the dominant one in our model.

\input{tablasX.tex}

\section{Summary and conclusions}
\label{sect::conclusions}

In this work we have studied the nature of the lowest-lying negative-parity
$\Lambda$ resonances with strange, charm or bottom flavors, with
$J^P=\frac{1}{2}^-$ and $\frac{3}{2}^-$. To this end we have adopted a
description based on pseudoscalar and vector mesons interacting in $s$-wave
with $\frac{1}{2}^+$ and $\frac{3}{2}^+$ baryons. The model, spelled out in
\cite{Garcia-Recio:2013gaa}, is based on spin-flavor and heavy-quark
extensions of the WT interaction, thereby embodying the correct symmetries in
the appropriate limits, such as chiral symmetry and HQSS. (The symmetries are
explicitly broken at the level of masses and decay constants of the basic
hadrons.)  The interaction is then used as an input of the Bethe-Salpeter
equation in coupled-channels. The model has no
free parameters, barring the choice of subtraction point in the
renormalization of the loop function. This is fixed by a prescription, or
occasionally used to modify the positions of the resonances to fulfill
phenomenological constraints.

As already uncovered by previous studies, we find a double pole structure for
the states with $J^P=\frac{1}{2}^-$ and a single pole for the states with
$\frac{3}{2}^-$ for each of the three flavors. The novelty comes from the
systematic study of the composition of these resonances, as a function
 of the heavy quark mass, addressing the question of to what extent the
structure of the resonances is fully saturated by the available
$s$-wave meson-baryon channels.

Regarding the overall compositeness of the nine $\Lambda$ resonances studied,
we find that for a given flavor sector, the closer to threshold (on the
complex plane) the better the resonance is described as an $s$-wave
meson-baryon molecule. Also, the heavier the flavor the higher the
compositeness $1-Z$.  More explicitly, we find that $1-Z$ is large for the first $\Lambda(\frac{1}{2}^-)$ of each
flavor and the compositeness decreases as we move to the second
$\Lambda(\frac{1}{2}^-)$ states and then to the $\Lambda(\frac{3}{2}^-)$
ones. Also, the compositeness is large for all bottom $\Lambda$ states. This
would indicate that the three first $\Lambda(\frac{1}{2}^-)$ and all
 the
rest of the bottom resonances considered are largely saturated, regarding their
composition, by $s$-wave meson-baryon channels. This would not be so for the
strange and charmed second $\Lambda(\frac{1}{2}^-)$ and strange and charmed
$\Lambda(\frac{3}{2}^-)$ resonances, which would require further components to
achieve the saturation of the sum rule.

With respect to the detailed composition of the states, we find that the
first $\Lambda(\frac{1}{2}^-)$ states of each flavor couple strongly to
pseudoscalar-$N$ and vector-$N$ channels. This is a manifestation of
spin-flavor symmetry between pseudoscalar and vector partners, and in
particular HQSS in the charm and bottom cases. For the $\Lambda(1405)$ this
implies that the pseudoscalar-$N$ channel, being lighter than the vector-$N$
one, almost saturates the compositeness of the state.  It is noteworthy that a
large weight of $\bar{K}N$ in the $\Lambda(1405)$ has been recently reported
from lattice QCD calculations \cite{Hall:2014uca}. The situation changes for
charm and bottom flavors where the two channels ($DN$ and $D^*N$ or $\bar{B}N$
and $\bar{B}^*N$) are relatively closer due to HQSS. In this case, the weights
of the channels follows more closely the trend of the couplings. (Although
with smaller weights, a similar pattern appears for the strange partners of
the mesons, $D_s\Lambda$, $D_s^*\Lambda$, etc., due to $\SU(3)-$light flavor symmetry.)  These two channels almost saturate the
composition of the first $\Lambda_c(2595)$ and $\Lambda_b(5912)$ states.

For the second $\Lambda(\frac{1}{2}^-)$ states, the main observation is its
sizable coupling to the lightest channels $\pi \Sigma$, $\pi \Sigma_c$ and
$\pi \Sigma_b$. As a consequence these states are effectively more excited
than the first $\Lambda(\frac{1}{2}^-)$ states and for the strange and charm
flavor this explains their larger widths, as compared to the first states. The
larger phase space also implies that higher partial waves could play a role,
consistently with the fact that they are much less saturated by the $s$-wave
meson-baryon channels considered here.

Another observation is the similar structure of the second
$\Lambda(\frac{1}{2}^-)$ and $\Lambda(\frac{3}{2}^-)$ states, which appear as
HQSS or spin-flavor partners. This is particularly clear in the bottom case,
where HQSS works better. The couplings of $\pi\Sigma_b$ in the second
$\Lambda_b(5912)$ and $\pi\Sigma_b^*$ in $\Lambda_b(5920)$ are similar and,
the same pattern is seen for $\bar{B}N$ and $\bar{B}^*N$.  This translates to
the corresponding composition weights, although distorted by the effect of
different excitation energy of the channels. The spin-flavor symmetry between
$\Sigma_c$ and $\Sigma_c^*$, and $\Sigma$ and $\Sigma^*$ still acts for charm
and strange flavors.

Although beyond the scope of the present work, it would be interesting to
consider also quark models and try to compare to hadronic results in order to
see whether the composition of a given resonance can be termed as molecular,
made of quarks or hybrid, and if possible to quantify the hybrid mixture.

\appendix

\section{Derivative of the loop function}
\label{app:Gp}

In order to compute analytically the derivative of the ($s$-wave) loop
function required in Eq.~(\ref{eq:sumruledef}), we recall its definition in
Eq.~(\ref{eq::funcion_loop}):
\begin{equation}
G = i ~2 M \int \,\frac{d^{4}q}{\left( 2\pi \right)^{4}} 
\, \frac{1}{q^{2} - m^2+i\epsilon}\,\, \frac{1}{\left(P-q\right)^{2} - M^2+i\epsilon} 
,\qquad
M,m>0
\,.
\end{equation}
Choosing the C.M. frame, $P^\mu=(\sqrt{s},\mathbf{0})$, 
its partial derivative with respect to the energy can be written as
\begin{equation}
G^\prime(\sqrt{s}) \equiv
\frac{\partial G}{\partial \sqrt{s}} =  
-i \, 4 M\int \frac{d^{4}q}{\left( 2\pi \right)^{4}}\, 
\,\frac{1}{q^{2} - m^2+i\epsilon}\,
\, \frac{P^0-q^0}{\left( \left(P-q\right)^{2} - M^2 + i\epsilon\right)^2}
.
\end{equation}
Unlike the loop function, $G^\prime(\sqrt{s})$ is ultraviolet
convergent.  The use of a
standard Feynman's parameterization (see e.g. Eq.~(10.13) of
\cite{Mandl:1985bg}) gives
\begin{equation}
G^\prime = 
-i \, 8 M  \int \frac{d^{4}q}{\left( 2\pi \right)^{4}} 
\int_0^1 dx \frac{x (P^0-q^0) }{\left[   (q - x P)^2 - x^2 P^2 - m^2 
+ (P^2 - M^2 + m^2)x + i\epsilon
\right]^3}  ,
\end{equation}
and after a translation in the integration variable:
\begin{equation}
G^\prime = -i\,8 M  \sqrt{s} \int_0^1 dx 
\int \frac{d^{4}q}{\left( 2\pi \right)^{4}}  
\frac{x(1-x)}{\left(   q^2 +x(1- x) s -(1-x) m^2 - x M^2 
 + i\epsilon
 \right)^3}  
.
\end{equation}
The integral over the $q^\mu$ is now straightforward (using e.g. Eq.~(A.44) of
\cite{Peskin:1995ev}) to obtain
\begin{equation}
G^\prime = \frac{M \sqrt{s}}{4\pi^2} \int_0^1 dx
\frac{ 1 }{ s -  \frac{m^2}{x} - \frac{ M^2}{1-x} + i\epsilon } 
.
\label{eq:A5}
\end{equation}
It follows that $G^\prime$ is purely real for $s<(M+m)^2$, while ${\rm Im}
G^\prime <0$ for $s>(M+m)^2$.

The integral Eq.~(\ref{eq:A5}) is well defined for $\sqrt{s}$ on the
complex plane, excluding $s = (M + m)^2$, and it yields
\begin{equation}
G^\prime = \frac{M}{4\pi^2 s\sqrt{s}}
\left(
s-(M^2-m^2)\log\frac{M}{m}-\frac{s(M^2+m^2)-(M^2-m^2)^2}{\sqrt{s-s_+}\sqrt{s-s_-}}
\log\frac{\sqrt{s-s_+}-\sqrt{s-s_-}}{\sqrt{s-s_+}+\sqrt{s-s_-}} 
\right)
\label{eq:dGdE}
\end{equation}
with
\begin{equation}
 s_{\pm} = (M\pm m)^2,
\qquad
\mathrm{Arg}\sqrt{s-s_+}\in[0,\pi),\quad 
\mathrm{Arg}\sqrt{s-s_-} \in \left[-\frac{\pi}{2},\frac{\pi}{2} \right[
, \quad
\mathrm{Im} \log \in[0,\pi]
\qquad \text{(FRS)}
.
\label{eq:FRS}
\end{equation}

The function $G^\prime(\sqrt{s})$ inherits the branching points and Riemann
sheet structure of the loop function $G$.  The expression in
Eqs. (\ref{eq:dGdE}) and (\ref{eq:FRS}) corresponds to the so-called First
Riemann Sheet (FRS) with respect to the $s_+$ branching point and the branch
cut is along $s \ge s_+ $. For the FRS, the point $s=s_{-}$ is a regular
point. $s_+$ is a branching point of order one (by circling twice around $s_+$
the function returns to its original value) hence there is a Second Riemann
Sheet (SRS) with respect to $s_+$ that continues the FRS at the two borders of
the selected cut. The SRS is obtained by analytic continuation. The point
$s=s_-$ is a branching point of order one in the SRS. However, for the
physical phase space of interest the new Riemann sheets introduced by the
branching at $s_-$ are not relevant. The expression of $G^\prime(\sqrt{s})$ on
the SRS takes the same form as in Eq.~(\ref{eq:dGdE}) but taking
\begin{equation}
\mathrm{Arg}\sqrt{s-s_+}\in[0,\pi),\qquad 
\mathrm{Arg}\sqrt{s-s_-} \in [\frac{\pi}{2},\frac{3\pi}{2} )
,\qquad
\mathrm{Im}\log \in[\pi,2\pi]
\qquad \text{(SRS)}.
\label{eq:SRS}
\end{equation}

\begin{figure*}
\begin{center}
\includegraphics[width=0.45\textwidth]{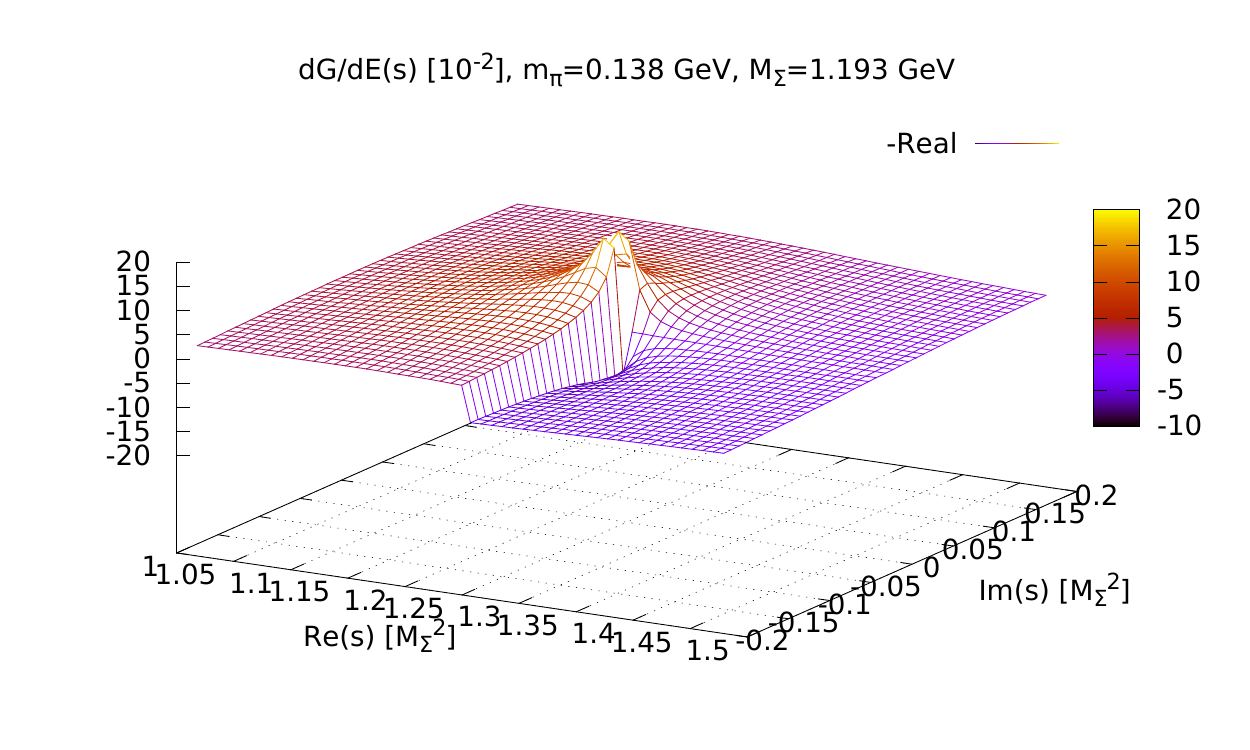}
\includegraphics[width=0.45\textwidth]{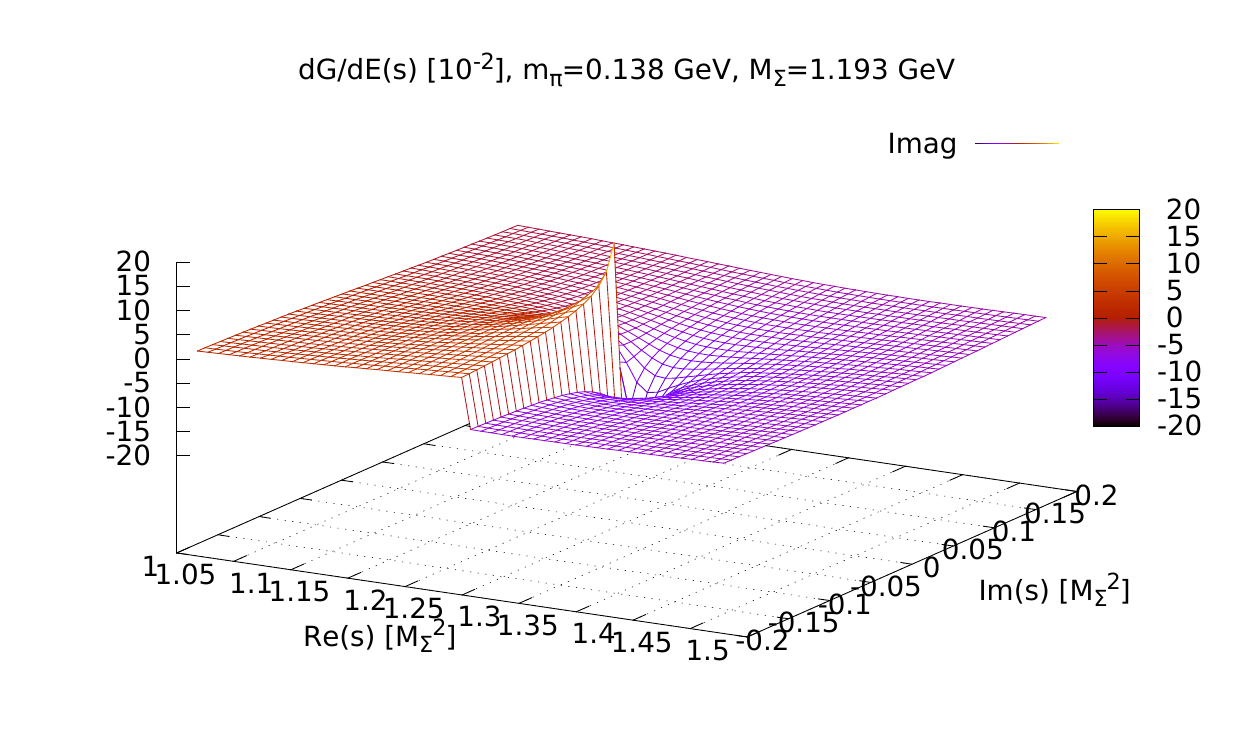}
\caption{\small Function $G^\prime(\sqrt{s})$. On the left $-{\rm
    Re}G^\prime$. On the right $+{\rm Im}G^\prime$.  The display corresponds to
  $m=m_\pi$ and $M= M_\Sigma$.}
\label{fig:2}
\end{center}
\end{figure*}

The function $G^\prime(\sqrt{s})$ is displayed in Fig.~{\ref{fig:2}}. In the
plot $\sqrt{s}$ is on the FRS when ${\rm Re}(s)<(M+m)^2$ or when ${\rm Re}(s)>(M+m)^2$
and ${\rm Im}(s)>0$, and on the SRS when ${\rm Re}(s)>(M+m)^2$ and
${\rm Im}(s)<0$. The bound states fall on the ``negative'' (with respect to
$M+m$) real axis and the resonances fall below the ``positive'' real axis, for
the relevant channel. This cut of the complex plane covers most cases. An
exception is the $\Lambda(1520)$ with $\sqrt{\alpha} = 0.780$, which falls
slightly at the left of the branch cut, for $\pi\Sigma^*$.

For completeness we give here analytical form of the loop function, subtracted at
$s=s_+$:
\begin{equation}
G = - \frac{M}{8\pi^2 s}
\left(
(s-s_+) \frac{M-m}{M+m} \log\frac{M}{m}
+ \sqrt{s-s_+}\sqrt{s-s_-} 
\log\frac{\sqrt{s-s_+}-\sqrt{s-s_-}}{\sqrt{s-s_+}+\sqrt{s-s_-}} 
\right)
.
\end{equation}
The choices of branches for the FRS and the SRS are as in Eqs. (\ref{eq:FRS}) and 
(\ref{eq:SRS}), respectively.

\section{Compositeness sum rule}
\label{app:sumrule}

In this appendix we prove the relation in
Eq.~(\ref{eq::sumrule_energy_dependent}).

We start from Eq.~(\ref{eq:bse})
\begin{equation}
T(\sqrt{s}) = (V^{-1}(\sqrt{s}) - G(\sqrt{s}))^{-1} 
\end{equation}
where $T$, $V$ and $G$ are matrices and $G$ is diagonal.  Taking
a derivative with respect to $\sqrt{s}$ (using the operator identity $\delta (A^{-1}) =
-A^{-1} \delta A A^{-1}$)
\begin{equation}
T^\prime =  T \left( G^\prime +
V^{-1} V^\prime V^{-1} \right) T
\,.
\label{eq:B2}
\end{equation}
On the other hand, Eq.~(\ref{Tfit}) implies
\begin{equation}
T_{ij} = \frac{g_i g_j}{\Delta}  + R_{ij}(\sqrt{s})
,\qquad
\Delta \equiv \sqrt{s}-\sqrt{s_R}
,
\label{eq:B3}
\end{equation}
where the remainder $R_{ij}$ is regular at the pole. Taking a derivative with
respect to $\Delta$, substituting $T$ in Eq.~(\ref{eq:B2}) and multiplying by
$\Delta^2$ gives, at $\Delta=0$,
\begin{equation}
- g_i g_j =  \sum_{k,l} g_i g_k \left( G_k^\prime \delta_{kl} +
\sum_{r,s} (V^{-1})_{kr} V^\prime_{rs} (V^{-1})_{sl} \right)_{\Delta=0} g_l g_j
\,.
\label{eq:B4}
\end{equation}
Since at least one of the couplings must be different from zero (to have a
pole) it follows that
\begin{equation}
-1  =  \sum_{k,l} g_k g_l \left( G_k^\prime \delta_{kl} +
\sum_{r,s} (V^{-1})_{kr} V^\prime_{rs} (V^{-1})_{sl} \right)\Bigg|_{\Delta=0}
\,.
\label{eq:B5}
\end{equation}
To arrive to Eq.~(\ref{eq::sumrule_energy_dependent})
it only remains to show that $(V^{-1})_{sl}$ can be replaced by $ \delta_{sl} G_l$
in Eq.~(\ref{eq:B5}). This follows from $V^{-1}=T^{-1}+G$ and  $\sum_l
(T^{-1})_{jl} g_l|_{\Delta=0}=0$. The latter equality can be deduced
 from,
\begin{equation}
\sum_{l} (T^{-1})_{jl} \left(\frac{g_l g_j}{\Delta}  + R_{lj}(\sqrt{s})\right) = 1 
\end{equation}
without summing over the index $j$, which trivially follows from Eq.~(\ref{eq:B3})
and $T^{-1} T =1 $. Thus, in
the limit $\Delta\to 0$, we find
\begin{equation}
\lim_{\Delta \to 0}\sum_l (T^{-1})_{jl} g_l =
\lim_{\Delta \to 0} \frac{\Delta}{g_j} = 0\,.
\end{equation}
The statement is trivial if there is just one channel.
More generally, the singular part of $T$ in Eq.~(\ref{eq:B3}) is a matrix of rank
one, the corresponding one-dimensional subspace being spanned by the vector
$g_i$, in coupled-channels space. The combination $\sum_l (T^{-1})_{sl} g_l$
selects that subspace, in which $T^{-1}$ vanishes at the pole.\footnote{The
  argument assumes that, at the pole, $T$ is a regular matrix on the complementary subspace. For
  instance, it would not work for
 $T= \begin{pmatrix} \frac{g^2}{\Delta} & 1 \\ 1 & 0\end{pmatrix} $, however,
    such a singular complement requires $V$ to be singular at the resonance
    pole. We assume that this is not the case since $V^\prime$ would not exist
    at the pole.}

\section*{Acknowledgments}
We thank E. Oset for  careful proof reading and comments. 
C.~H.-D. thanks the support of the JAE-CSIC Program.
This research was supported by  Spanish Ministerio de Econom\'\i a y Competitividad and
European FEDER funds  under
contracts FPA2010-16963, FIS2011-28853-C02-02, FPA2013-43425-P,
FIS2014-59386-P,  FIS2014-51948-C2-1-P and 
FIS2014-57026-REDT.  LT acknowledges support from the Ram\'on y
Cajal Research Programme from Ministerio de Ciencia e Innovaci\'on and from
FP7-PEOPLE-2011-CIG under Contract No. PCIG09-GA-2011-291679.

\end{document}

%% file: masses.tex
\begin{table*}

\begin{center}
\begin{tabular}{| cr r c  r c c | cr r r c c |}
\hline
Meson & mass & width & decay
& ~$\SU(6)_{ U_C(1)}$ & $\SU(3)_{ 2J+1}$ & HQSS 
& Baryon & mass & width
& ~$\SU(6)_{ U_C(1)}$ & $\SU(3)_{ 2J+1}$ & HQSS
\\
&&& constant &&&& &&&&&\cr
\hline
$\pi$ & $138.04$ &  & $\phantom{0}92.4$
& $\bm{35_{0}}$ & $\bm{8_1}$ & singlet 
& $N$ & $938.92$ &  
& $\bm{56_{0}}$ & $\bm{8_2}$ & singlet
\\

$K$ & $495.65$  &  & $113.0$
& $\bm{35_{0}}$ & $\bm{8_1}$ & singlet 
& $\Lambda$ & $1115.68$ &  
& $\bm{56_{0}}$ & $\bm{8_2}$ & singlet
\\

$\eta$ & $547.86$ &   & $111.0$
& $\bm{35_{0}}$ & $\bm{8_1}$ & singlet 
& $\Sigma$ & $1193.15$ &  
& $\bm{56_{0}}$ & $\bm{8_2}$ & singlet
\\

$\rho$ & $775.49$ & 150  & $153.0$
& $\bm{35_{0}}$ & $\bm{8_3}$ & singlet
& $\Xi$ & $1318.29$  &  
& $\bm{56_{0}}$ & $\bm{8_2}$ & singlet
\\

$K^*$ & $893.88$ & 50  & $153.0$
& $\bm{35_{0}}$ & $\bm{8_3}$ & singlet 
& $\Sigma^*$ & $1382.80$ & 35 
& $\bm{56_{0}}$ & $\bm{10_4}$ & singlet
\\

$\omega$ & $782.65$ &   & $138.0$
& $\bm{35_{0}}$ & ideal & singlet
& $\Xi^*$ & $1531.80$ &  
& $\bm{56_{0}}$ & $\bm{10_4}$ & singlet
\\

$\phi$ & $1019.46$ &   & $163.0$
& $\bm{35_{0}}$ & ideal & singlet
& $\Lambda_c$ & $2286.46$ &  
& $\bm{21_{1}}$ & $\bm{3^*_2}$ & singlet
\\

$\eta^\prime$ & $957.78$ &   & $111.0$ 
& $\bm{1_{0}}$ & $\bm{1_1}$ & singlet 
& $\Xi_c$ & $2469.34$ &    
& $\bm{21_{1}}$ & $\bm{3^*_2}$ & singlet
\\

$D$  & $1867.23$  &    & $157.4$
& $\bm{6^*_{1}}$ & $\bm{3^*_1}$ & doublet
& $\Sigma_c$ & $2453.54$ &  
& $\bm{21_{1}}$ & $\bm{6_2}$ & doublet
\\

$D^*$ & $2008.61$  &  & $f_D$
& $\bm{6^*_{1}}$ & $\bm{3^*_3}$ & doublet
& $ \Sigma_c^*$ & $2518.07$ &  
& $\bm{21_{1}}$ & $\bm{6_4}$ & doublet
\\

$D_s$ & $1968.30$  &  & $193.7$
& $\bm{6^*_{1}}$ & $\bm{3^*_1}$ & doublet
& $\Xi^\prime_c$ & $2576.75$ &  
& $\bm{21_{1}}$ & $\bm{6_2}$ & doublet
\\

$D_s^*$ & $2112.10$  &  & $f_{D_s}$
& $\bm{6^*_{1}}$ & $\bm{3^*_3}$ & doublet
& $\Xi_c^*$ & $2645.90$ &  
& $\bm{21_{1}}$ & $\bm{6_4}$ & doublet
\\

$B$  & $5279.42$  &  & $133.6$
& $\bm{6^*_{1}}$ & $\bm{3^*_1}$ & doublet
& $\Lambda_b$ &  $5619.50$ &  
& $\bm{21_{1}}$ & $\bm{3_2}$ & singlet
\\

$B^*$ &  $5325.20$   &  & $f_B$
& $\bm{6^*_{1}}$ & $\bm{3^*_3}$ & doublet
& $\Xi_b$ & $5794.00$ &  
& $\bm{21_{1}}$ & $\bm{3^*_2}$ & singlet
\\

$B_s$ &  $5366.77$  &  & $159.1$
& $\bm{6^*_{1}}$ & $\bm{3^*_1}$ & doublet
& $ \Sigma_b$ & $5813.40$ &  
& $\bm{21_{1}}$ & $\bm{6_2}$ & doublet
\\

$B_s^*$ & $5415.40$  &  & $f_{B_s}$
& $\bm{6^*_{1}}$ & $\bm{3^*_3}$ & doublet
& $ \Sigma_b^*$ & $5833.60$ &  
& $\bm{21_{1}}$ & $\bm{6_4}$ & doublet
\\
&&&&&&
& $\Xi^\prime_b$ & $5926.00$ &  
& $\bm{21_{1}}$ & $\bm{6_2}$ & doublet
\\ 
&&&&&&
& $\Xi_b^*$ & $5949.30$ &  
& $\bm{21_{1}}$ & $\bm{6_4}$ & doublet \\
\hline
\end{tabular}
\end{center}
\caption{Baryon masses, $M_i$, meson masses, $m_i$, and meson decay
  constants, $f_i$, (in MeV) used throughout this work.  The widths in
  MeV units,
  $\Gamma_R$, used in the convolutions (Eq.~(\ref{eq:convolution})) are also
  provided. The masses and decay constants are taken from
  Refs.~\cite{GarciaRecio:2008dp,GarciaRecio:2012db}.  The
  $\SU(6)\times\SU_C(2)\times\U_C(1)$ and $\SU(3)\times \SU(2)$ labels are
  displayed as well  (for simplicity we do not explicitly give the
  spin of the heavy quark sector, since it is trivially 0 or 1/2.) The last column indicates the HQSS multiplets. Members of
  a doublet are placed in consecutive rows. }
\label{tab:masses}
\end{table*}

%% file: t00m112.tex
\begin{table*}[b!]
\caption{
  Matrix elements $D_{ij}$ for the  $\Lambda(1405)$ sector:
$C=B=0,~S=-1,~I=0,~J^P=1/2^-$.
}
\label{tab:00m002}
\begin{tabular}{r|rrrrrrrrrrrrrrrrrrrrrrrrrrrrrrrrrrrrrrrrrrrrrrrr}
 &  $ \Sigma \pion $ &  $ \Nucleon 
  \kaonb $ &  $ \Lambda \eta $ 
 &  $ \Cascada \kaon $ &  $ \Nucleon 
  \kaonSb $ &  $ \Lambda \omega $ 
 &  $ \Sigma \rho $ &  $ \Lambda \phi 
  $ &  $ \SigmaS \rho $ 
 &  $ \Cascada \kaonS $ &  $ \CascadaS 
  \kaonS $\\
\hline
$ \Sigma \pion $& $ -4 $ 
 & $ \sqrt{\frac{ 3 }{ 2 }} 
   $ & $ 0 $ 
 & $ -\sqrt{\frac{ 3 }{ 2 }} 
   $ & $ \sqrt{\frac{ 1 }{ 2 
    }} $ & $ 0 $ 
 & $ \sqrt{\frac{ 64 }{ 3 }} 
   $ & $ 0 $ 
 & $ \sqrt{\frac{ 32 }{ 3 }} 
   $ & $ \sqrt{\frac{ 25 }{ 2 
    }} $ & $ 2 $ \\
$ \Nucleon \kaonb $
 & $ \sqrt{\frac{ 3 }{ 2 }} 
   $ & $ -3 $ 
 & $ -\sqrt{\frac{ 9 }{ 2 }} 
   $ & $ 0 $ 
 & $ \sqrt{ 27 } $ 
 & $ \sqrt{\frac{ 9 }{ 2 }} 
   $ & $ \sqrt{\frac{ 1 }{ 2 
    }} $ & $ 3 $ & $ 2 $ 
 & $ 0 $ & $ 0 $ \\
$ \Lambda \eta $& $ 0 $ 
 & $ -\sqrt{\frac{ 9 }{ 2 }} 
   $ & $ 0 $ 
 & $ \sqrt{\frac{ 9 }{ 2 }} 
   $ & $ \sqrt{\frac{ 27 }{ 2 
    }} $ & $ 0 $ & $ 0 $ 
 & $ 0 $ & $ 0 $ 
 & $ -\sqrt{\frac{ 3 }{ 2 }} 
   $ & $ \sqrt{ 12 } $ \\
$ \Cascada \kaon $
 & $ -\sqrt{\frac{ 3 }{ 2 }} 
   $ & $ 0 $ 
 & $ \sqrt{\frac{ 9 }{ 2 }} 
   $ & $ -3 $ & $ 0 $ 
 & $ -\sqrt{\frac{ 1 }{ 2 }} 
   $ & $ \sqrt{\frac{ 25 }{ 2 
    }} $ & $ -1 $ 
 & $ -2 $ 
 & $ \sqrt{ 3 } $ 
 & $ 0 $ \\
$ \Nucleon \kaonSb $
 & $ \sqrt{\frac{ 1 }{ 2 }} 
   $ & $ \sqrt{ 27 } $ 
 & $ \sqrt{\frac{ 27 }{ 2 }} 
   $ & $ 0 $ & $ -9 $ 
 & $ \sqrt{\frac{ 3 }{ 2 }} 
   $ & $ \sqrt{\frac{ 25 }{ 6 
    }} $ & $ -\sqrt{ 27 } 
   $ & $ -\sqrt{\frac{ 4 }{ 3 
     }} $ & $ 0 $ 
 & $ 0 $ \\
$ \Lambda \omega $& $ 0 $ 
 & $ \sqrt{\frac{ 9 }{ 2 }} 
   $ & $ 0 $ 
 & $ -\sqrt{\frac{ 1 }{ 2 }} 
   $ & $ \sqrt{\frac{ 3 }{ 2 
    }} $ & $ 0 $ & $ 4 $ 
 & $ 0 $ & $ \sqrt{ 8 } 
   $ & $ \sqrt{\frac{ 25 }{ 6 
    }} $ & $ \sqrt{\frac{ 4 
    }{ 3 }} $ \\
$ \Sigma \rho $
 & $ \sqrt{\frac{ 64 }{ 3 }} 
   $ & $ \sqrt{\frac{ 1 }{ 2 
    }} $ & $ 0 $ 
 & $ \sqrt{\frac{ 25 }{ 2 }} 
   $ & $ \sqrt{\frac{ 25 }{ 6 
    }} $ & $ 4 $ 
 & $ -\frac{ 20 }{ 3  } $ 
 & $ 0 $ 
 & $ \sqrt{\frac{ 8 }{ 9 }} 
   $ & $ -\sqrt{\frac{ 169 }{ 6 
     }} $ & $ \sqrt{\frac{ 4 
    }{ 3 }} $ \\
$ \Lambda \phi $& $ 0 $ 
 & $ 3 $ & $ 0 $ 
 & $ -1 $ 
 & $ -\sqrt{ 27 } $ 
 & $ 0 $ & $ 0 $ 
 & $ -4 $ & $ 0 $ 
 & $ \sqrt{\frac{ 1 }{ 3 }} 
   $ & $ -\sqrt{\frac{ 8 }{ 3 
     }} $ \\
$ \SigmaS \rho $
 & $ \sqrt{\frac{ 32 }{ 3 }} 
   $ & $ 2 $ & $ 0 $ 
 & $ -2 $ 
 & $ -\sqrt{\frac{ 4 }{ 3 }} 
   $ & $ \sqrt{ 8 } $ 
 & $ \sqrt{\frac{ 8 }{ 9 }} 
   $ & $ 0 $ 
 & $ -\frac{ 22 }{ 3  } $ 
 & $ -\sqrt{\frac{ 4 }{ 3 }} 
   $ & $ -\sqrt{\frac{ 128 }{ 3 
     }} $ \\
$ \Cascada \kaonS $
 & $ \sqrt{\frac{ 25 }{ 2 }} 
   $ & $ 0 $ 
 & $ -\sqrt{\frac{ 3 }{ 2 }} 
   $ & $ \sqrt{ 3 } $ 
 & $ 0 $ 
 & $ \sqrt{\frac{ 25 }{ 6 }} 
   $ & $ -\sqrt{\frac{ 169 }{ 6 
     }} $ & $ \sqrt{\frac{ 1 
    }{ 3 }} $ 
 & $ -\sqrt{\frac{ 4 }{ 3 }} 
   $ & $ -\frac{ 19 }{ 3  } 
   $ & $ -\sqrt{\frac{ 32 }{ 9 
     }} $ \\
$ \CascadaS \kaonS $& $ 2 $ 
 & $ 0 $ 
 & $ \sqrt{ 12 } $ 
 & $ 0 $ & $ 0 $ 
 & $ \sqrt{\frac{ 4 }{ 3 }} 
   $ & $ \sqrt{\frac{ 4 }{ 3 
    }} $ 
 & $ -\sqrt{\frac{ 8 }{ 3 }} 
   $ & $ -\sqrt{\frac{ 128 }{ 3 
     }} $ 
 & $ -\sqrt{\frac{ 32 }{ 9 }} 
   $ & $ -\frac{ 14 }{ 3  } 
   $ \\
\end{tabular}
\end{table*}

%% file: tablasX.tex
\begin{table*}[b!]
\caption{Calculated masses, widths and compositeness of the negative-parity
  $\Lambda$ states in the strange sector. The coupling constants and the
  weights of the various channels are also displayed. The main numbers refer
  to the default value $\alpha=1$, while the numbers in parenthesis refer to
  the same quantities computed with a subtraction point chosen so that the
  masses are close to the experimental ones \cite{Agashe:2014kda}. For this
  purpose similar masses have been adopted for the two $\Lambda(\frac{1}{2}^-)$
  states. For each $\Lambda$ state the largest compositeness weights have been
  highlighted with boldface.}
\label{tab:La1520}
\begin{tabular}{|cccccc|crrrr|}
\hline
State & $J^P$ & $\sqrt{\alpha}$ & $M_R$ & $\Gamma_R$ & $1-Z$ 
& Channel & $|g_i|$~ & $g_i$~~~~~~ & $X_i$~~ &
$(X_i)$~~ \\
\hline
$\mathbf {\Lambda(1405)}$ & $\frac{1}{2}^- $ & $1$ & $ 1430.0 $ & $ 5.5 $ & $0.887$
& $\pi  \Sigma     $ &  $0.50$ & $ 0.19 + 0.46i $ & $-0.008$ & $(-0.006)$
\\
&& $(0.867)$ & $( 1405.1)$ & $ ( 12.8) $ & $(  0.772)$ &
 $ \bf{ \bar{K} N }$ &  $1.78$ & $-1.76 - 0.24i $ & $ \bf{0.795}$ & $ \bf{( 0.597)}$ 
\\
&&&&&&
 $\eta \Lambda $ &  $0.94$ & $ -0.94 - 0.06i $ & $ 0.023$ & $(0.040)$ 
\\
&&&&&&
 $K \Xi $ &  $0.10$ & $ 0.05 + 0.09i $ & $-0.000$ &$(-0.000)$ 
\\
&&&&&&
 $\bf{\bar{K}^* N} $ &  $2.18$ & $2.18 + 0.06i $ & $ \bf{0.066}$ & $\bf{(0.126)}$
\\
&&&&&&
 $\omega  \Lambda    $ &  $0.48$ & $ 0.48 + 0.06i $ & $ 0.003$ & $(0.004)$ \\
&&&&&&
 $\rho \Sigma     $ &  $0.34$ & $0.18 - 0.29i $ & $-0.001$ & $(-0.003)$ \\
&&&&&&
 $\phi \Lambda    $ &  $0.88$ & $ 0.88 + 0.05i $ & $ 0.007$ & $( 0.013)$ \\
&&&&&&
 $\rho \Sigma^*   $ &  $0.40$ & $ 0.39 - 0.09i $ & $ 0.002$ & $( 0.002)$ \\
&&&&&&
 $K^*  \Xi        $ &  $0.17$ & $0.03 - 0.17i $ & $-0.000$ & $(-0.001)$ \\
&&&&&&
 $K^*  \Xi^*      $ &  $0.06$ & $ 0.05 - 0.03i $ & $ 0.000$ & $( 0.000)$ \\
\\
\hline\hline

$\mathbf{\Lambda(1405)}$ & $\frac{1}{2}^- $ & $1$ &  $1373.0$ &  $170.0$ &  $ 0.332$ &
 $\bf{ \pi  \Sigma } $ & $ 2.62$ & $ 2.07 - 1.60i $ & $ \bf{0.353}$ & $\bf{( 0.586)}$ \\
&& $(1.12)$ & $( 1405.0)$ & $( 376.1)$ & $(  0.522)$ & 
 $\bar{K}    N    $ &  $1.03$ & $-0.78 + 0.67i $ & $-0.024$ & $(-0.018)$ \\
&&&&&&
 $\eta \Lambda    $ & $ 0.2$1 & $0.07 + 0.20i $ & $-0.001$ & $( 0.001)$  \\
&&&&&&
 $K    \Xi        $ &  $0.46$ & $ 0.29 - 0.36i $ & $-0.001$ & $(-0.005)$  \\
&&&&&&
 $\bar{K}^*    N  $ &  $0.57$ & $ -0.51 + 0.26i $ & $ 0.003$ & $( 0.004)$  \\
&&&&&&
 $\omega  \Lambda $ &  $0.24$ & $-0.02 - 0.24i $ & $-0.001$ & $( 0.002)$  \\
&&&&&&
 $\rho \Sigma     $ &  $1.68$ & $ -1.37 + 0.97i $ & $ 0.006$ & $(-0.023)$  \\
&&&&&&
 $\phi \Lambda    $ &  $0.17$ & $-0.07 - 0.16i $ & $-0.000$ & $( 0.000)$  \\
&&&&&&
 $\rho \Sigma^*   $ &  $0.66$ & $-0.45 + 0.48i $ & $-0.001$ & $(-0.011)$  \\
&&&&&&
 $K^*  \Xi        $ &  $0.93$ & $ -0.63 + 0.69i $ & $-0.002$ & $(-0.011)$  \\
&&&&&&
 $K^*  \Xi^*      $ &  $0.29$ & $ 0.01 + 0.29i $ & $-0.001$ & $(-0.003)$  \\

\\
\hline\hline

 $\mathbf{\Lambda(1520)}$ & $\frac{3}{2}^- $ & $ 1$ &$  1540.0$ &  $ 74.0 $ &  $0.274$ &
 $ \bf{ \pi\Sigma^*} $ &  $2.29$ & $ 2.03 - 1.05i $ & $ \bf{0.227}$ & $ \bf{( 0.109)}$ \\
&& $(0.780)$ & $ ( 1522.7) $ & $ ( 25.9) $ & $(  0.134)$ & 
 $\bar{K}^*  N $ & $ 0.87$ & $ 0.84 - 0.23i $ & $ 0.011$ & $( 0.005)$ \\
&&&&&&
 $\omega  \Lambda $ & $ 0.40$ & $-0.31 + 0.26i $ & $ 0.000$ & $( 0.001)$  \\
&&&&&&
 $\rho \Sigma     $ & $ 1.18$ & $ 1.09 - 0.46i $ & $ 0.015$ & $( 0.008)$  \\
&&&&&&
 $K    \Xi^*      $ & $ 0.63$ & $ 0.49 - 0.40i $ & $ 0.001$ & $( 0.001)$  \\
&&&&&&
 $\phi \Lambda    $ & $ 0.04$ & $0.02 - 0.03i $ & $ 0.000$ & $( 0.000)$  \\
&&&&&&
 $\rho \Sigma^*   $ &  $1.60$ & $ 1.42 - 0.74i $ & $ 0.015$ & $( 0.009)$  \\
&&&&&&
 $K^*  \Xi        $ &  $0.55$ & $-0.50 + 0.24i $ & $ 0.002$ & $( 0.001)$  \\
&&&&&&
 $K^*  \Xi^*      $ & $ 0.59$ & $ 0.46 - 0.37i $ & $ 0.001$ & $( 0.000)$  \\

\hline\hline

\end{tabular}
\end{table*}

\begin{table*}[b!]
\caption{Same as Table \ref{tab:La1520} for the charm sector.}
\label{tab:LaC2625}
\begin{tabular}{|cccccc|crrrr|}
\hline
State & $J^P$ & $\sqrt{\alpha}$ & $M_R$ & $\Gamma_R$ & $1-Z$ & Channel & $|g_i|$~ & $g_i$~~~~~~ & $X_i$~~ &
$(X^\prime_i)$~~ 
\\ 
\hline
$\mathbf{\Lambda_c(2595)}$  & $\frac{1}{2}^- $ & $1$ &  $2619.0$ &   $ 1.2 $  & $0.878$ & 
$\pi   \Sigma_c   $ &  $0.31$ & $ 0.22 + 0.22i $ & $-0.012$ & $(-0.023)$ \\
&& $(0.979)$ &$ ( 2592.3) $ &  $(   0.3) $ & $(  0.844)$ & 
$\bf{D N} $ &  $3.49$ & $-3.49 - 0.14i $ & $ \bf{0.275}$ & $\bf{(0.292)}$ \\
&&&&&&
 $\eta \Lambda_c $ & $ 0.40$ & $ 0.40 - 0.00i $ & $ 0.007$ & $( 0.009)$  \\
&&&&&&
 $ \bf{D^* N} $ &  $5.64$ & $-5.64 + 0.14i $ & $ \bf{0.465}$ & $\bf{(0.451)}$  \\
&&&&&&
 $K     \Xi_c  $ &  $0.22$ & $ 0.22 - 0.00i $ & $ 0.002$ & $( 0.001)$  \\
&&&&&&
$\omega\Lambda_c  $ &  $0.18$ & $ 0.18 + 0.04i $ & $ 0.001$ & $( 0.001)$  \\
&&&&&&
$K     \Xi'_c $ & $ 0.04$ & $ 0.02 + 0.04i $ & $-0.000$ & $( 0.000)$  \\
&&&&&&
 $D_s   \Lambda$ & $ 1.38$ & $-1.38 + 0.01i $ & $ 0.026$ & $( 0.026)$  \\
&&&&&&
 $\bf{D_s^* \Lambda}$ &  $2.87$ & $-2.87 + 0.03i $ & $ \bf{0.086}$ & $\bf{(0.057)}$  \\
&&&&&&
 $\rho  \Sigma_c   $ & $ 0.41$ & $ 0.39 + 0.12i $ & $ 0.003$ & $( 0.005)$  \\
&&&&&&
 $\eta' \Lambda_c  $ &  $0.92$ & $ 0.92 + 0.01i $ & $ 0.018$ & $( 0.018)$  \\
&&&&&&
 $\rho  \Sigma_c^* $ &  $0.58$ & $ 0.58 - 0.07i $ & $ 0.007$ & $( 0.006)$  \\
&&&&&&
 $\phi  \Lambda_c  $ & $ 0.0$1 & $ 0.01 + 0.00i $ & $ 0.000$ & $( 0.000)$  \\
&&&&&&
 $K^*   \Xi_c  $ &  $0.05$ & $ 0.02 + 0.05i $ & $ 0.000$ & $( 0.000)$  \\
&&&&&&
 $K^*   \Xi'_c $ &  $0.16$ & $ 0.16 + 0.04i $ & $ 0.000$ & $( 0.000)$  \\
&&&&&&
 $K^*   \Xi_c^*$ & $ 0.15$ & $ 0.15 + 0.02i $ & $ 0.000$ & $( 0.000)$  \\
\\
\hline\hline

$\mathbf{\Lambda_c(2595)}$  & $\frac{1}{2}^- $ & $1$ & $2617.0$ &   $90.0$ &  $ 0.401$  &
 $\bf{\pi \Sigma_c} $ & $ 2.36$ & $ 2.09 - 1.09i $ & $ \bf{0.325}$ & $\bf{(0.252)}$ \\
&& (0.950) & $( 2595.0)$ & $(  36.8)$ & $(  0.354)$ &
 $D           N      $ & $ 1.64$ & $-1.46 + 0.75i $ & $ 0.027$ & $(0.015)$  \\
&&&&&&
 $\eta  \Lambda_c  $ & $ 0.06$ & $ 0.02 - 0.06i $ & $-0.000$ & $( 0.000)$  \\
&&&&&&
 $\bf{D^* N}$ &  $1.43$ & $ 1.34 + 0.51i $ & $ 0.024$ & $\bf{(0.057)}$  \\
&&&&&&
 $K     \Xi_c  $ &  $0.04$ & $ 0.02 - 0.03i $ & $ 0.000$ & $( 0.000)$  \\
&&&&&&
 $\omega\Lambda_c  $ & $ 0.43$ & $ 0.30 - 0.31i $ & $ 0.000$ & $( 0.003)$  \\
&&&&&&
 $K     \Xi'_c $ & $ 0.48$ & $ 0.38 - 0.29i $ & $ 0.001$ & $( 0.002)$  \\
&&&&&&
 $D_s   \Lambda $ & $ 0.21$ & $ 0.07 + 0.20i $ & $ 0.000$ & $( 0.001)$  \\
&&&&&&
 $D_s^* \Lambda $ & $ 0.40$ & $ 0.22 + 0.33i $ & $-0.001$ & $( 0.002)$  \\
&&&&&&
 $\rho  \Sigma_c   $ &  $1.28$ & $ 1.11 - 0.63i $ & $ 0.016$ & $( 0.013)$  \\
&&&&&&
 $\eta' \Lambda_c  $ &  $0.13$ & $-0.07 - 0.11i $ & $-0.000$ & $( 0.001)$  \\
&&&&&&
 $\rho  \Sigma_c^* $ & $ 0.70$ & $-0.64 + 0.28i $ & $ 0.006$ & $( 0.006)$  \\
&&&&&&
 $\phi  \Lambda_c  $ &  $0.01$ & $ 0.01 + 0.01i $ & $ 0.000$ & $( 0.000)$  \\
&&&&&&
 $K^*   \Xi_c  $ &  $0.51$ & $ 0.45 - 0.25i $ & $ 0.002$ & $( 0.002)$  \\
&&&&&&
 $K^*   \Xi'_c $ & $ 0.29$ & $ 0.10 - 0.27i $ & $ 0.000$ & $( 0.001)$  \\
&&&&&&
 $K^*   \Xi_c^* $ & $ 0.20$ & $-0.15 + 0.13i $ & $ 0.000$ & $( 0.000)$  \\
\\
\hline\hline

$\mathbf{\Lambda_c(2625)}$  & $\frac{3}{2}^- $ & $1$ & $2667.0$ &   $55.0$ & $0.365$ &
 $\bf{\pi \Sigma_c^*} $ &  $2.19$ & $ 1.97 - 0.95i $ & $ \bf{0.268}$ & $\bf{(0.319)}$ \\
&& $(0.985)$ & $( 2628.1)$ & $(   0.0)$ & $(  0.405)$  & 
$\bf{ D^* N }$ &  $2.03$ & $ 1.96 - 0.51i $ & $ \bf{0.057}$ & $\bf{(0.044)}$ \\
&&&&&&
 $\omega \Lambda_c  $ & $ 0.53$ & $-0.45 + 0.28i $ & $ 0.003$ & $( 0.018)$  \\
&&&&&&
 $K     \Xi_c^* $ & $ 0.42$ & $ 0.34 - 0.24i $ & $ 0.002$ & $( 0.001)$  \\
&&&&&&
 $D_s^* \Lambda $ &  $0.06$ & $ 0.05 - 0.04i $ & $ 0.000$ & $( 0.000)$  \\
&&&&&&
 $\rho  \Sigma_c   $ &  $0.75$ & $ 0.68 - 0.31i $ & $ 0.008$ & $( 0.005)$  \\
&&&&&&
 $\rho  \Sigma_c^* $ &  $1.30$ & $ 1.17 - 0.57i $ & $ 0.022$ & $( 0.013)$  \\
&&&&&&
 $\phi  \Lambda_c  $ &  $0.0$1 & $-0.01 + 0.01i $ & $ 0.000$ & $( 0.000)$  \\
&&&&&&
 $K^*   \Xi_c  $ &  $0.61$ & $-0.55 + 0.27i $ & $ 0.005$ & $( 0.004)$  \\
&&&&&&
 $K^*   \Xi'_c $ &  $0.25$ & $ 0.22 - 0.12i $ & $ 0.001$ & $( 0.000)$  \\
&&&&&&
 $K^*   \Xi_c^* $ &  $0.40$ & $ 0.33 - 0.23i $ & $ 0.001$ & $( 0.001)$  \\

\hline\hline

\end{tabular}
\end{table*}

\begin{table*}[b!]
\caption{Same as Table \ref{tab:La1520} for the beauty sector.}
\label{tab:LaB5920}
\begin{tabular}{|cccccc|crrr|}
\hline
State & $J^P$ & $\sqrt{\alpha}$ & $M_R$ & $\Gamma_R$ & $1-Z$ & Channel & $g_i$~ & $X_i$~~ &
$(X^\prime_i)$~~
\\ 
\hline

$\mathbf{\Lambda_b(5912)}$  & $\frac{1}{2}^- $ & 1 & $5878.0$ &   $ 0.0$ & $ 0.956$ &
 $\pi   \Sigma_b   $ & $ 0.04  $ & $ 0.000$ & $( 0.000)$ \\
&& (1.01) & $( 5912.1)$ & $(   0.0)$ & $(  0.958)$ &
 $\bf{ \bar{B} N}$  & $-4.55  $ & $ \bf{0.205}$ & $\bf{(0.217)}$ \\
&&&&&&
 $\eta  \Lambda_b  $  & $ 0.33  $ & $ 0.006$ & $( 0.010)$  \\
&&&&&&
 $ \bf{ \bar{B^*} N} $  & $-7.70  $ & $ \bf{0.539}$ & $ \bf{(0.561)}$  \\
&&&&&&
 $K     \Xi_b  $  & $ 0.22  $ & $ 0.002$ & $( 0.002)$  \\
&&&&&&
 $\omega\Lambda_b  $  & $ 0.04  $ & $ 0.000$ & $( 0.000)$  \\
&&&&&&
 $K     \Xi'_b $  & $ 0.02  $ & $ 0.000$ & $( 0.000)$  \\
&&&&&&
 $\bar{B_s}\Lambda $  & $-1.96  $ & $ 0.031$ & $( 0.031)$  \\
&&&&&&
$\bf{ \bar{B_s^*} \Lambda} $  & $-4.01  $ & $ \bf{0.122}$ & $ \bf{(0.084)}$  \\
&&&&&&
 $\rho  \Sigma_b   $  & $ 0.38  $ & $ 0.005$ & $( 0.006)$  \\
&&&&&&
 $\eta' \Lambda_b  $  & $ 0.96  $ & $ 0.032$ & $( 0.032)$  \\
&&&&&&
 $\rho  \Sigma_b^* $  & $ 0.57  $ & $ 0.011$ & $( 0.013)$  \\
&&&&&&
 $\phi  \Lambda_b  $  & $ 0.02  $ & $ 0.000$ & $( 0.000)$  \\
&&&&&&
 $K^*   \Xi_b  $  & $-0.01  $ & $ 0.000$ & $( 0.000)$  \\
&&&&&&
 $K^*   \Xi'_b $  & $ 0.17  $ & $ 0.001$ & $( 0.000)$  \\
&&&&&&
 $K^*   \Xi_b^* $  & $ 0.19  $ & $ 0.001$ & $( 0.001)$  \\
\\
\hline\hline

$\mathbf{\Lambda_b(5912)}$ & $\frac{1}{2}^- $ & 1 & $5949.0$ &   $ 0.0$ &  $0.865$ &
 $ \bf{\pi \Sigma_b }$  & $ 1.31 $ & $ \bf{0.698}$ & $ \bf{(0.397)}$ \\
&& (0.984) & $( 5912.0)$ & $(   0.0)$ & $(  0.788)$ &
 $ \bf{ \bar{B} N} $  & $-2.90 $ & $ \bf{0.096}$ & $ \bf{(0.215)}$ \\
&&&&&&
 $\eta  \Lambda_b  $  & $ 0.01 $ & $ 0.000$ & $( 0.000)$  \\
&&&&&&
 $ \bf{\bar{B^*} N} $  & $ 1.91 $ & $ 0.038$ & $\bf{( 0.082)}$  \\
&&&&&&
 $K     \Xi_b  $  & $-0.01 $ & $ 0.000$ & $( 0.000)$  \\
&&&&&&
 $ \bf{ \omega\Lambda_b}  $ & $ 0.78 $ & $ 0.028$ & $\bf{(0.088)}$  \\
&&&&&&
 $K     \Xi'_b $  & $ 0.18 $ & $ 0.001$ & $( 0.001)$  \\
&&&&&&
 $\bar{B_s}\Lambda $  & $-0.01 $ & $ 0.000$ & $( 0.000)$  \\
 &&&&&&
$\bar{B_s^*} \Lambda $  & $ 0.18 $ & $ 0.000$ & $( 0.000)$  \\
 &&&&&&
$\rho  \Sigma_b   $  & $ 0.13 $ & $ 0.001$ & $( 0.002)$  \\
&&&&&&
 $\eta' \Lambda_b  $  & $-0.03 $ & $ 0.000$ & $( 0.000)$  \\
&&&&&&
 $\rho  \Sigma_b^* $  & $-0.08 $ & $ 0.000$ & $( 0.001)$  \\
 &&&&&&
$\phi  \Lambda_b  $  & $-0.00 $ & $0.000$ & $( 0.000)$  \\
&&&&&&
 $K^*   \Xi_b  $  & $ 0.23 $ & $ 0.002$ & $( 0.002)$  \\
&&&&&&
 $K^*   \Xi'_b $  & $ 0.13 $ & $ 0.001$ & $( 0.000)$  \\
&&&&&&
 $K^*   \Xi_b^* $  & $-0.10 $ & $ 0.000$ & $( 0.000)$  \\
\\
\hline\hline

 $\mathbf{\Lambda_b(5920)}$ & $\frac{3}{2}^- $ & 1 & $5963.0$ &   $ 0.0 $ &  $ 0.818$ &
 $ \bf{ \pi \Sigma_b^*} $  & $ 1.54  $ & $ \bf{0.581}$ & $\bf{(0.356)}$ \\
&& (0.983) & $ ( 5919.7) $ & $ (   0.0) $ &  $(  0.785)$ &
 $ \bf{ \bar{B^*} N }$  & $ 4.16  $ & $ \bf{0.185}$ & $\bf{(0.319)}$ \\
&&&&&&
 $ \bf{ \omega\Lambda_b } $  & $-0.99  $ & $ 0.046 $ & $\bf{(0.102)}$  \\
&&&&&&
 $K     \Xi_b^* $  & $ 0.20  $ & $ 0.002$ & $( 0.001)$  \\
&&&&&&
 $\bar{B_s^*} \Lambda $  & $ 0.14  $ & $ 0.000$ & $( 0.000)$  \\
&&&&&&
 $\rho  \Sigma_b   $  & $ 0.08  $ & $ 0.000$ & $( 0.001)$  \\
&&&&&&
 $\rho  \Sigma_b^* $  & $ 0.12  $ & $ 0.001$ & $( 0.005)$  \\
&&&&&&
 $\phi  \Lambda_b  $  & $ 0.00  $ & $ 0.000$ & $( 0.000)$  \\
&&&&&&
 $K^*   \Xi_b  $  & $-0.28  $ & $ 0.003$ & $( 0.002)$  \\
&&&&&&
 $K^*   \Xi'_b $  & $ 0.08  $ & $ 0.000$ & $( 0.000)$  \\
&&&&&&
 $K^*   \Xi_b^* $  & $ 0.17  $ & $ 0.001$ & $( 0.000)$  \\

\hline\hline

\end{tabular}
\end{table*}

%% file: paper.bbl
\begin{thebibliography}{0}


\bibitem{cleo} http://w4.lns.cornell.edu/Research/CLEO/

\bibitem{belle} http://belle.kek.jp/

\bibitem{bes} http://bes3.ihep.ac.cn/

\bibitem{babar} http://www-public.slac.stanford.edu/babar/

\bibitem{panda} http://www-panda.gsi.de/

\bibitem{jparc} http://j-parc.jp/index-e.html

\bibitem{Aaij:2012da} 
  R.~Aaij {\it et al.}  [LHCb Collaboration],
  Phys.\ Rev.\ Lett.\  {\bf 109}, 172003 (2012)
  

\bibitem{Tolos:2004yg}
  L.~Tolos, J.~Schaffner-Bielich and A.~Mishra,
  Phys.\ Rev.\  C {\bf 70}, 025203 (2004).

\bibitem{Tolos:2005ft} 
  L.~Tolos, J.~Schaffner-Bielich and H.~Stoecker,
  Phys.\ Lett.\ B {\bf 635}, 85 (2006)
  
\bibitem{Lutz:2003jw}
  M.~F.~M.~Lutz and E.~E.~Kolomeitsev,
  Nucl.\ Phys.\  A {\bf 730}, 110 (2004).

\bibitem{Lutz:2005ip} 
  M.~F.~M.~Lutz and E.~E.~Kolomeitsev,
  Nucl.\ Phys.\ A {\bf 755}, 29 (2005)
  
\bibitem{Hofmann:2005sw} 
  J.~Hofmann and M.~F.~M.~Lutz,
  Nucl.\ Phys.\ A {\bf 763}, 90 (2005)

\bibitem{Hofmann:2006qx} 
  J.~Hofmann and M.~F.~M.~Lutz,
  Nucl.\ Phys.\ A {\bf 776}, 17 (2006)


\bibitem{korpa-lutz}
  M.~F.~M.~Lutz and C.~L.~Korpa,
  Phys.\ Lett.\  B {\bf 633}, 43 (2006).


\bibitem{Mizutani:2006vq} 
  T.~Mizutani and A.~Ramos,
  Phys.\ Rev.\ C {\bf 74}, 065201 (2006)

\bibitem{Tolos:2007vh}
  L.~Tolos, A.~Ramos and T.~Mizutani,
  Phys.\ Rev.\  C {\bf 77}, 015207 (2008).


\bibitem{JimenezTejero:2009vq} 
  C.~E.~Jimenez-Tejero, A.~Ramos and I.~Vidana,
  Phys.\ Rev.\ C {\bf 80}, 055206 (2009)

\bibitem{JimenezTejero:2011fc} 
  C.~E.~Jimenez-Tejero, A.~Ramos, L.~Tolos and I.~Vidana,
  Phys.\ Rev.\ C {\bf 84}, 015208 (2011)
 
\bibitem{Haidenbauer:2007jq}
  J.~Haidenbauer, G.~Krein, U.~G.~Meissner and A.~Sibirtsev,
  Eur.\ Phys.\ J.\  A {\bf 33}, 107 (2007).

\bibitem{Haidenbauer:2008ff}
  J.~Haidenbauer, G.~Krein, U.~G.~Meissner and A.~Sibirtsev,
  Eur.\ Phys.\ J.\  A {\bf 37}, 55 (2008).

\bibitem{Haidenbauer:2010ch}
  J.~Haidenbauer, G.~Krein, U.~G.~Meissner and L.~Tolos,
  Eur. Phys. J A {\bf 47}, 18 (2011)

\bibitem{Wu:2010jy}
  J.~-J.~Wu, R.~Molina, E.~Oset and B.~S.~Zou,
  Phys.\ Rev.\ Lett.\  {\bf 105}, 232001 (2010);

\bibitem{Wu:2010vk}
  J.~-J.~Wu, R.~Molina, E.~Oset and B.~S.~Zou,
  Phys.\ Rev.\ C {\bf 84} (2011) 015202

\bibitem{Oset2012} 
  E.~Oset, A.~Ramos, E.~J.~Garzon, R.~Molina, L.~Tolos, C.~W.~Xiao, J.~J.~Wu and B.~S.~Zou,
  International Journal of Modern Physics E, Vol.\  {\bf 21}, 1230011 (2012)

 
\bibitem{Liang:2014kra} 
  W.~H.~Liang, T.~Uchino, C.~W.~Xiao and E.~Oset,
  Eur.\ Phys.\ J.\ A {\bf 51}, no. 2, 16 (2015)
  
\bibitem{Liang:2014eba} 
  W.~H.~Liang, C.~W.~Xiao and E.~Oset,
  Phys.\ Rev.\ D {\bf 89}, no. 5, 054023 (2014)

\bibitem{Wu:2010rv} 
  J.~-J.~Wu and B.~S.~Zou,
  Phys.\ Lett.\ B {\bf 709}, 70 (2012) 



\bibitem{IW89} N. Isgur and M.B. Wise, Phys. Lett. B {\bf 232},  113 (1989).

\bibitem{Ne94}  M. Neubert, Phys. Rep. {\bf 245},  259 (1994).

\bibitem{MW00} A.V. Manohar and M.B. Wise, {\it Heavy Quark Physics},
  Cambridge Monographs on Particle Physics, Nuclear Physics and
  Cosmology, vol. 10



\bibitem{GarciaRecio:2008dp} 
  C.~Garcia-Recio, V.~K.~Magas, T.~Mizutani, J.~Nieves, A.~Ramos, L.~L.~Salcedo and L.~Tolos,
  Phys.\ Rev.\ D {\bf 79}, 054004 (2009)
  
\bibitem{Gamermann:2010zz} 
  D.~Gamermann, C.~Garcia-Recio, J.~Nieves, L.~L.~Salcedo and L.~Tolos,
  Phys.\ Rev.\ D {\bf 81}, 094016 (2010)

\bibitem{Romanets:2012hm} 
  O.~Romanets, L.~Tolos, C.~Garcia-Recio, J.~Nieves, L.~L.~Salcedo and R.~G.~E.~Timmermans,
  Phys.\ Rev.\ D {\bf 85}, 114032 (2012)
  
\bibitem{GarciaRecio:2012db} 
  C.~Garcia-Recio, J.~Nieves, O.~Romanets, L.~L.~Salcedo and L.~Tolos,
  Phys.\ Rev.\ D {\bf 87}, no. 3, 034032 (2013)
  
\bibitem{Garcia-Recio:2013gaa} 
  C.~Garcia-Recio, J.~Nieves, O.~Romanets, L.~L.~Salcedo and L.~Tolos,
  Phys.\ Rev.\ D {\bf 87}, 074034 (2013)
  
\bibitem{Tolos:2013gta} 
  L.~Tolos,
  Int.\ J.\ Mod.\ Phys.\ E {\bf 22}, 1330027 (2013)

\bibitem{Tolos:2009nn} 
  L.~Tolos, C.~Garcia-Recio and J.~Nieves,
  Phys.\ Rev.\ C {\bf 80}, 065202 (2009)

\bibitem{GarciaRecio:2010vt} 
  C.~Garcia-Recio, J.~Nieves and L.~Tolos,
  Phys.\ Lett.\ B {\bf 690}, 369 (2010)


\bibitem{GarciaRecio:2011xt} 
  C.~Garcia-Recio, J.~Nieves, L.~L.~Salcedo and L.~Tolos,
  Phys.\ Rev.\ C {\bf 85}, 025203 (2012)

\bibitem{Xiao:2013yca} 
  C.~W.~Xiao, J.~Nieves and E.~Oset,
  Phys.\ Rev.\ D {\bf 88}, 056012 (2013)

\bibitem{Ozpineci:2013zas} 
  A.~Ozpineci, C.~W.~Xiao and E.~Oset,
  Phys.\ Rev.\ D {\bf 88}, 034018 (2013)


\bibitem{Weinberg:1962hj} 
  S.~Weinberg,
  Phys.\ Rev.\  {\bf 130}, 776 (1963).


\bibitem{Weinberg:1965zz} 
  S.~Weinberg,
  Phys.\ Rev.\  {\bf 137}, B672 (1965).

\bibitem{Hanhart:2010wh} 
  C.~Hanhart, Y.~S.~Kalashnikova and A.~V.~Nefediev,
  Phys.\ Rev.\ D {\bf 81}, 094028 (2010)


\bibitem{Baru:2003qq} 
  V.~Baru, J.~Haidenbauer, C.~Hanhart, Y.~Kalashnikova and A.~E.~Kudryavtsev,
  Phys.\ Lett.\ B {\bf 586}, 53 (2004)
  
\bibitem{Cleven:2011gp} 
  M.~Cleven, F.~K.~Guo, C.~Hanhart and U.~G.~Meissner,
  Eur.\ Phys.\ J.\ A {\bf 47}, 120 (2011)
  
  
\bibitem{Gamermann:2009uq} 
  D.~Gamermann, J.~Nieves, E.~Oset and E.~Ruiz Arriola,
  Phys.\ Rev.\ D {\bf 81}, 014029 (2010)
  
  
\bibitem{YamagataSekihara:2010pj} 
  J.~Yamagata-Sekihara, J.~Nieves and E.~Oset,
  Phys.\ Rev.\ D {\bf 83}, 014003 (2011)
  
\bibitem{Aceti:2012dd} 
  F.~Aceti and E.~Oset,
  Phys.\ Rev.\ D {\bf 86}, 014012 (2012)

\bibitem{Xiao:2012vv} 
  C.~W.~Xiao, F.~Aceti and M.~Bayar,
  Eur.\ Phys.\ J.\ A {\bf 49}, 22 (2013)

\bibitem{Aceti:2014ala} 
  F.~Aceti, L.~R.~Dai, L.~S.~Geng, E.~Oset and Y.~Zhang,
  Eur.\ Phys.\ J.\ A {\bf 50}, 57 (2014)

\bibitem{Hyodo:2011qc} 
  T.~Hyodo, D.~Jido and A.~Hosaka,
  Phys.\ Rev.\ C {\bf 85}, 015201 (2012)

\bibitem{Nagahiro:2014mba} 
  H.~Nagahiro and A.~Hosaka,
  Phys.\ Rev.\ C {\bf 90}, no. 6, 065201 (2014)

\bibitem{Aceti:2014wka} 
  F.~Aceti, E.~Oset and L.~Roca,
  Phys.\ Rev.\ C {\bf 90}, no. 2, 025208 (2014)
  
\bibitem{Sekihara:2012xp} 
  T.~Sekihara and T.~Hyodo,
  Phys.\ Rev.\ C {\bf 87}, no. 4, 045202 (2013)

\bibitem{Sekihara:2014kya} 
  T.~Sekihara, T.~Hyodo and D.~Jido,
  arXiv:1411.2308 [hep-ph].

\bibitem{Hyodo:2013nka} 
  T.~Hyodo,
  Int.\ J.\ Mod.\ Phys.\ A {\bf 28}, 1330045 (2013)





\bibitem{GarciaRecio:2005hy} 
  C.~Garcia-Recio, J.~Nieves and L.~L.~Salcedo,
  Phys.\ Rev.\ D {\bf 74}, 034025 (2006)
  
\bibitem{GarciaRecio:2006wb} 
  C.~Garcia-Recio, J.~Nieves and L.~L.~Salcedo,
  Phys.\ Rev.\ D {\bf 74}, 036004 (2006)
  
\bibitem{Toki:2007ab} 
  H.~Toki, C.~Garcia-Recio and J.~Nieves,
  Phys.\ Rev.\ D {\bf 77}, 034001 (2008)
  
\bibitem{GarciaRecio:2010ki} 
  C.~Garcia-Recio, L.~S.~Geng, J.~Nieves and L.~L.~Salcedo,
  Phys.\ Rev.\ D {\bf 83}, 016007 (2011)
  
\bibitem{Gamermann:2011mq} 
  D.~Gamermann, C.~Garcia-Recio, J.~Nieves and L.~L.~Salcedo,
  Phys.\ Rev.\ D {\bf 84}, 056017 (2011)
  
\bibitem{Garcia-Recio:2013uva} 
  C.~Garcia-Recio, L.~S.~Geng, J.~Nieves, L.~L.~Salcedo, E.~Wang and J.~J.~Xie,
  Phys.\ Rev.\ D {\bf 87}, no. 9, 096006 (2013)
  
\bibitem{GarciaRecio:2010vf} 
  C.~Garcia-Recio and L.~L.~Salcedo,
  J.\ Math.\ Phys.\  {\bf 52}, 043503 (2011)

\bibitem{Nieves:2001wt} 
  J.~Nieves and E.~Ruiz Arriola,
  Phys.\ Rev.\ D {\bf 64}, 116008 (2001)

\bibitem{Roca:2006sz} 
  L.~Roca, S.~Sarkar, V.~K.~Magas and E.~Oset,
  Phys.\ Rev.\ C {\bf 73}, 045208 (2006)
  
\bibitem{Gamermann:2007fi} 
  D.~Gamermann and E.~Oset,
  Eur.\ Phys.\ J.\ A {\bf 33}, 119 (2007)
    
    
\bibitem{GarciaRecio:2003ks} 
  C.~Garcia-Recio, M.~F.~M.~Lutz and J.~Nieves,
  Phys.\ Lett.\ B {\bf 582}, 49 (2004)
  
\bibitem{Hyodo:2008xr} 
  T.~Hyodo, D.~Jido and A.~Hosaka,
  Phys.\ Rev.\ C {\bf 78}, 025203 (2008)

   

\bibitem{Jido:2003cb} 
  D.~Jido, J.~A.~Oller, E.~Oset, A.~Ramos and U.~G.~Meissner,
  Nucl.\ Phys.\ A {\bf 725}, 181 (2003)
  
\bibitem{GarciaRecio:2002td} 
  C.~Garcia-Recio, J.~Nieves, E.~Ruiz Arriola and M.~J.~Vicente Vacas,
  Phys.\ Rev.\ D {\bf 67}, 076009 (2003)

\bibitem{Hall:2014uca} 
  J.~M.~M.~Hall, W.~Kamleh, D.~B.~Leinweber, B.~J.~Menadue, B.~J.~Owen, A.~W.~Thomas and R.~D.~Young,
  Phys.\ Rev.\ Lett.\  {\bf 114}, no. 13, 132002 (2015)




\bibitem{Agashe:2014kda} 
  K.~A.~Olive {\it et al.}  [Particle Data Group Collaboration],
  Chin.\ Phys.\ C {\bf 38}, 090001 (2014).

\bibitem{Mandl:1985bg} 
  F.~Mandl and G.~Shaw,
  Chichester, Uk: Wiley (2010) 478 p.

\bibitem{Peskin:1995ev} 
  M.~E.~Peskin and D.~V.~Schroeder,
  Reading, USA: Addison-Wesley (1995) 842 p

\end{thebibliography}
